\documentclass[11pt,a4paper]{article}
\newcommand{\keywords}[1]{\vspace{2ex}\noindent\textbf{Keywords:} #1\vspace{2ex}}
\usepackage[a4paper,margin=1in]{geometry}

\usepackage{amsmath}
\usepackage{amssymb}
\usepackage{amsfonts}
\usepackage{amsthm}
\usepackage{mathtools}
\usepackage{physics}
\usepackage{bm}
\usepackage{titling}

\usepackage{graphicx}
\usepackage{booktabs}
\usepackage{multirow}
\usepackage{float}

\usepackage[numbers,sort&compress]{natbib}

\usepackage[colorlinks=true,
            linkcolor=blue,
            citecolor=blue,
            urlcolor=blue]{hyperref}

\usepackage{xcolor}
\usepackage{enumitem}
\usepackage{setspace}
\usepackage{titlesec}

\title{\Large\bfseries
Revealing the Quantum Signature of Gravity via Gravitational Waves}

\author{%
\textbf{Partha Nandi}$^{1,2}$\\[0.5em]
$^{1}$Department of Physics, Stellenbosch University, Stellenbosch 7600, South Africa\\
$^{2}$National Institute for Theoretical and Computational Sciences (NITheCS), South Africa\\[0.5em]
\texttt{pnandi@sun.ac.za}
}

\date{}

\begin{document}

\maketitle

%\begin{abstract}
%...
%\end{abstract}


\bigskip

%%%%%%%%%%%%%%%%%%%%%%%%%%%%%%%%%%%%%%%%%%%%%%%%%%%%%%%%%%%%%%%
\begin{abstract}

Can propagating gravitational waves serve as operational probes of the
quantum nature of gravity? We address this question by developing a
unified theoretical framework that combines spacetime geometry, quantum
information, and gravitational-wave physics. Starting from the geodesic
deviation equation in linearized General Relativity, we derive the
effective detector Hamiltonian directly from spacetime geometry and
construct the complete quantum dynamics for detector subsystems
interacting with both classical and quantized propagating
gravitational-wave fields. This unified formulation enables a direct
comparison between classical and quantum descriptions of gravitational
radiation within the same physical framework. We demonstrate that
classical gravitational-wave backgrounds can induce mixedness in the
detector state but cannot generate genuine quantum correlations between
the detector subsystems. In contrast, quantized gravitational waves
coherently mediate gravity-induced entanglement, quantum coherence,
quantum memory, and nonclassical correlations, providing clear
operational signatures of the quantum nature of propagating
gravitational radiation. We further discuss how mesoscopic quantum
mechanical oscillators offer a promising route towards experimentally
probing these effects. Our results establish a geometric and
quantum-information-based framework for exploring quantum gravity
through propagating gravitational waves.

\end{abstract}

%%%%%%%%%%%%%%%%%%%%%%%%%%%%%%%%%%%%%%%%%%%%%%%%%%%%%%%%%%%%%%%


%%%%%%%%%%%%%%%%%%%%%%%%%%%%%%%%%%%%%%%%%%%%%%%%%%%%%%%%%%%%%%%

\section{Introduction}

The direct detection of gravitational waves by the LIGO--Virgo
Collaboration opened a new era in observational astronomy and provided one
of the most remarkable confirmations of Einstein's General Theory of
Relativity in the dynamical strong-field regime~\cite{abbott}. Since then,
gravitational-wave observations have become a powerful probe of compact
astrophysical objects, the dynamics of spacetime, and the evolution of the
Universe~\cite{Haskell:2023yrv}. These observations have firmly established
gravitational waves as observable dynamical degrees of freedom of the
classical gravitational field. They also motivate a more fundamental
question: \emph{can propagating gravitational waves themselves reveal the
quantum nature of gravity?}

Understanding whether gravity is fundamentally classical or quantum remains
one of the central unresolved problems in modern physics. Considerable
progress has been achieved through quantum field theory in curved spacetime
and through candidate theories of quantum gravity~\cite{Kiefer2004},
including string theory~\cite{Green:1987sp}, loop quantum gravity
\cite{RovelliVidotto2014}, causal dynamical triangulations, asymptotic
safety, and noncommutative geometry~\cite{Nandi:2023tfq,Scholtz:2025gfj}.
Nevertheless, direct experimental evidence for quantum gravitational
degrees of freedom remains lacking~\cite{PhysRevLett.129.131301}. A major
obstacle is the extreme weakness of gravitational interactions at the
quantum scale, which makes direct detection of individual gravitons and
other microscopic quantum gravitational effects extraordinarily difficult.
This has motivated the search for indirect operational signatures of the
quantum nature of gravity.

A particularly influential approach was proposed independently by Bose
\emph{et al.}~\cite{Bose} and Marletto and Vedral
\cite{marletto,Marletto:2017kzi}, who showed that gravity-mediated
entanglement between initially independent quantum systems can serve as an
operational probe of the mediator. The underlying idea is that, under
appropriate assumptions, a purely classical mediator cannot generate
entanglement between quantum systems through local interactions and
classical communication~\cite{Chitambar:2014svb, RickPerche:2025guz}.
Consequently, the observation of entanglement generated solely through a
gravitational interaction can provide evidence that the mediating degrees
of freedom cannot be described as purely classical. These proposals have
stimulated substantial theoretical and experimental activity, with the
original proposals focusing on weak-field gravitational interactions in
the Newtonian regime~\cite{Bose,marletto,Marletto:2017kzi}

A natural question is whether the same operational idea can be extended to
the genuinely dynamical regime of propagating gravitational waves. In
contrast to the static Newtonian gravitational interaction, gravitational
waves represent propagating degrees of freedom of spacetime. If such a field
is quantized, it constitutes a dynamical quantum subsystem that can in
principle become correlated with localized quantum systems. This raises the
possibility that two spatially separated quantum detectors, each coupled
locally to the same propagating gravitational field, could become entangled
without any direct interaction between the detectors themselves. Whether
such gravity-mediated entanglement can arise from propagating gravitational
radiation, and how it differs from the correlations produced by a classical
stochastic gravitational wave, is the central question addressed in this
work.

Current gravitational-wave detectors are not designed to address this
question. Modern laser interferometers operate in a regime in which quantum
fluctuations are dominated by the optical measurement system
\cite{PhysRevLett.121.031101,Jia:2024iqe}, while the test masses are
effectively described as classical objects. Consequently, the standard
description of an observed gravitational wave does not provide a direct
operational distinction between a classical gravitational-wave field and a
quantized gravitational field. A different class of detector, possessing
genuinely quantum mechanical degrees of freedom and capable of maintaining
coherence during its interaction with the gravitational field, is therefore
required to investigate this question.

In this work, we develop such a framework using localized quantum
oscillators coupled to a weak propagating gravitational wave. Starting from
the geodesic deviation equation in linearized General Relativity
\cite{PhysRevD.51.1701}, we derive the effective detector--gravitational-field
interaction directly from the local tidal response of freely falling test
masses. This construction provides a common geometric starting point for
both classical and quantized descriptions of the gravitational wave,
allowing the corresponding detector dynamics to be compared within the same
physical framework.

A central feature of our construction is the strict locality of the
detector--field interaction. Each detector couples only to the gravitational
field evaluated at its own spacetime position, and no direct interaction is
introduced between the detector subsystems. Any correlation generated
between the detectors must therefore arise through their common interaction
with the propagating gravitational field. This setup realizes the essential
mediator structure underlying gravity-mediated-entanglement proposals while
extending it from static gravitational interactions to propagating
gravitational radiation. In particular, the gravitational field is treated
as an independent propagating degree of freedom rather than as a field
generated by the detector subsystems themselves.

The quantum description also makes it possible to distinguish detector
mixedness from genuine detector--detector entanglement. Interaction with the
gravitational field can correlate the detectors with the gravitational
degrees of freedom and thereby reduce the purity of the detector subsystem
after the gravitational field is traced out. Such mixedness, however, is
not by itself a signature of quantum gravity, since classical stochastic
gravitational waves can also produce statistical mixtures of detector
states. The relevant distinction is whether the reduced detector state
develops nonseparable quantum correlations. We therefore analyze the
reduced detector state using the positive-partial-transpose criterion and
the negativity, while using the von Neumann entropy and purity to
characterize its mixedness \cite{Dutta:2025bge}.

Within this framework, we find that the classical and quantized
descriptions of the same propagating gravitational-wave field lead to
qualitatively different detector correlations. For a classical stochastic
field, the local detector dynamics remains separable and stochastic
averaging produces a convex mixture of product states. In contrast, when
the gravitational-wave field is quantized, the field acts as a common
quantum mediator and can generate entanglement between the two detector
subsystems. The leading contribution originates from the coherent
interference of the two indistinguishable excitation pathways associated
with the two detectors. Thus, the resulting detector entanglement is not a
consequence of any direct detector--detector coupling, but arises from their
local interaction with the same quantized propagating gravitational field.

Our analysis is intended as an operational study of propagating
gravitational radiation rather than as a complete theory of quantum
gravity. The gravitational field is treated in the weak-field regime of
linearized General Relativity, and the detector--field interaction is
analyzed perturbatively. We employ a single-mode description to isolate
the leading dynamical mechanism. Extensions to multimode gravitational
fields, more general initial field states, and a more realistic treatment
of the gravitational environment \cite{Nandi:2026sww} are natural
directions for future work.

Finally, we discuss possible detector platforms capable of realizing the
required quantum degrees of freedom. Mesoscopic quantum mechanical
oscillators are particularly attractive because of their large zero-point
motion, long coherence times, and high degree of quantum control
\cite{Aspelmeyer2014}. Although the extremely weak gravitational coupling
makes an immediate experimental observation challenging, such systems
provide a possible route toward probing quantum signatures of propagating
gravitational radiation that are inaccessible to conventional
kilometer-scale laser interferometers.

The remainder of this paper is organized as follows. In Sec.~II, we discuss the limitations of present gravitational-wave detectors for probing the quantum nature of gravity and motivate the use of quantum detector degrees of freedom. In Sec.~III, we introduce the general information-transfer paradigm underlying the interaction between quantum systems and physical fields. In Sec.~IV, we derive the effective detector Hamiltonian directly from the geodesic deviation equation in linearized General Relativity. In Sec.~V, we formulate the system Hamiltonian and the interaction with a quantized gravitational-wave mode, and develop the corresponding perturbative detector--graviton dynamics. In Sec.~VI, we investigate gravity-induced entanglement and characterize the reduced detector state for both classical stochastic and quantized gravitational fields, including the resulting entanglement negativity, von Neumann entropy, and purity. In Sec.~VII, we discuss the physical interpretation of our results, their relation to other approaches to gravity-mediated quantum correlations, the limitations of the single-mode and perturbative approximations, and possible experimental prospects. Finally, Sec.~VIII presents our conclusions.

%%%%%%%%%%%%%%%%%%%%%%%%%%%%%%%%%%%%%%%%%%%%%%%%%%%%%%%%%%%%%%%
\section{Why Go Beyond Classical Gravitational-Wave Detectors?}
%%%%%%%%%%%%%%%%%%%%%%%%%%%%%%%%%%%%%%%%%%%%%%%%%%%%%%%%%%%%%%%

The direct detection of gravitational waves by the
LIGO--Virgo--KAGRA Collaboration has established laser interferometry as
one of the most sensitive experimental techniques ever developed,
reaching strain sensitivities of order
$10^{-23}/\sqrt{\mathrm{Hz}}$
over the most sensitive frequency bands of current ground-based
interferometers
~\cite{abbott,aggiore2008,VIRGO:2014yos,PhysRevD.88.043007}.
These remarkable achievements have not only provided stringent tests of
General Relativity, but have also established gravitational-wave astronomy
as a new observational window into compact astrophysical systems and the
dynamical Universe. Nevertheless, present gravitational-wave observatories
are designed primarily to measure classical perturbations of spacetime.
They therefore establish the existence and properties of gravitational
waves without directly addressing the more fundamental question of whether
the gravitational field itself possesses quantum degrees of freedom.

The operating principle of a laser interferometer is conceptually
straightforward. A passing gravitational wave perturbs the spacetime
metric and produces a differential variation in the proper lengths of the
two orthogonal interferometer arms. This differential displacement is
converted into an optical phase shift of coherent laser light, from which
the gravitational-wave strain is reconstructed. The measured phase shift
may be written as
\begin{equation}
\Delta\phi
=
\frac{4\pi}{\lambda}
\Delta L_{\rm opt},
\label{phase}
\end{equation}
where $\lambda$ is the laser wavelength and $\Delta L_{\rm opt}$ denotes
the optical path-length difference induced by the gravitational wave.

Although the gravitational-wave signal is described within classical
General Relativity, the optical measurement process is quantum mechanical.
The discrete nature of light produces photon-number fluctuations and
associated phase uncertainty, commonly referred to as shot noise. The
corresponding displacement uncertainty can be expressed as
\begin{equation}
\Delta x_{\rm meas}
=
\frac{\lambda}{4\pi}
\Delta\phi_{\rm shot},
\label{shotnoise}
\end{equation}
which limits the precision with which the mirror displacement can be
determined~\cite{Caves1981,Kimble2001}.

Increasing the laser power reduces the relative contribution of shot noise
by improving the statistical precision of the optical phase measurement.
At the same time, however, the increased photon flux enhances
radiation-pressure fluctuations and consequently induces stochastic
motion of the interferometer mirrors. The competition between measurement
imprecision and radiation-pressure backaction gives rise to the familiar
Standard Quantum Limit (SQL) for continuous position measurements
\cite{Braginsky1995,Caves1981,Kimble2001,Clerk2010}.

It is important to distinguish these quantum measurement effects from
quantum fluctuations of the gravitational field itself. The SQL arises
from the quantum fluctuations of the optical measurement apparatus:
shot noise originates from the particle nature of light, while
radiation-pressure noise results from fluctuations in the momentum
transferred by the photons to the mirrors. In the standard description,
the gravitational wave remains a prescribed classical solution of
Einstein's equations throughout the measurement. Thus, present
interferometers perform quantum-limited measurements of a classical
spacetime signal rather than direct measurements of the quantum state of
the gravitational field.

The interferometer mirrors are also effectively classical test masses for
the purpose of gravitational-wave detection. Although the mirrors are
macroscopic quantum systems, quantum fluctuations of the interferometer
can induce mirror displacements at the level of $\sim 10^{-20},\mathrm{m}$,
as demonstrated experimentally for the $40,\mathrm{kg}$ test masses of
Advanced LIGO~\cite{LIGOScientific:2020luc}. By comparison, astrophysical gravitational
waves can produce differential arm-length changes of order
$10^{-18},\mathrm{m}$ in kilometer-scale interferometers. Thus, we may
characterize the relevant displacement scales as
\begin{equation}
\Delta x_{\rm zpf}
\sim
10^{-20},\mathrm{m}
\ll
\Delta L_{\rm GW}
\sim
10^{-18},\mathrm{m},
\label{zpf}
\end{equation}
where $\Delta x_{\rm zpf}$ denotes the characteristic quantum
displacement scale of the mirrors arising from their zero-point and
quantum back-action fluctuations, while $\Delta L_{\rm GW}$ denotes the
gravitational-wave-induced differential arm-length change. Consequently,
the mirror trajectories can be treated to an excellent approximation as
classical test-mass trajectories, while the quantum nature of the optical
field enters through the measurement process and its associated quantum
back-action.

The measured interferometric signal may therefore be schematically
decomposed as
\begin{equation}
\Delta L_{\rm opt}
=
\Delta L_{\rm GW}
+
\Delta x_{\rm rad}
+
\Delta x_{\rm meas},
\label{signal}
\end{equation}
where $\Delta L_{\rm GW}$ is the gravitational-wave-induced displacement,
$\Delta x_{\rm rad}$ represents mirror motion generated by
radiation-pressure fluctuations, and $\Delta x_{\rm meas}$ denotes the
measurement uncertainty associated with photon shot noise.

This distinction is central to the motivation of the present work.
Although quantum mechanics limits the sensitivity of current
gravitational-wave detectors, the relevant quantum fluctuations originate
from the detector and its optical readout rather than from the gravitational
field itself. Consequently, the conventional interferometric measurement
scheme does not provide an operational distinction between a classical
gravitational wave and a quantized gravitational field.

To investigate quantum properties of the gravitational field, the role of
the detector must therefore be changed. Rather than functioning as an
effectively classical test mass whose displacement is measured by an
optical readout, the detector should itself remain a coherent quantum
system during its interaction with the gravitational field. In this regime,
the gravitational field and detector constitute a joint quantum system, and
the interaction can generate quantum correlations that are inaccessible
through a purely classical displacement measurement.

Recent developments in cavity optomechanics, optomechanical crystals,
levitated nanoparticles, trapped ions, superconducting electromechanical
systems, and other mesoscopic platforms have opened access to mechanical
systems in which quantum fluctuations of the motional degrees of freedom
can be controlled and measured with increasing precision
\cite{Aspelmeyer2014,Clerk2010,RevModPhys.89.035002,Schnabel:2022evt}.
Compared with the kilogram-scale test masses used in kilometer-scale
interferometers, mesoscopic quantum systems can exhibit comparatively
large zero-point motion, high mechanical quality factors, and a high
degree of quantum control. These properties make them promising
candidates for investigating quantum aspects of gravitational
interactions~\cite{LIGOScientific:2020luc,PhysRevLett.124.221102}.

Mesoscopic mechanical oscillators can have effective masses many orders
of magnitude below those of kilogram-scale interferometer test masses,
while retaining sufficiently high mechanical quality factors and quantum
control~\cite{Aspelmeyer2014,PhysRevLett.124.221102}. For a representative
mesoscopic oscillator with effective mass
\begin{equation}
m\sim10^{-12}-10^{-9}\,\mathrm{kg},
\end{equation}
the zero-point motion can be substantially larger than that of
kilogram-scale interferometer test masses for comparable mechanical
frequencies. This motivates considering a quantum-sensitive regime
characterized by the condition
\begin{equation}
\Delta x_{\rm zpf}
\gtrsim
\Delta x_{\rm GW},
\label{quantumregime}
\end{equation}
where $\Delta x_{\rm GW}$ denotes the gravitational-wave-induced
displacement. Equation~\eqref{quantumregime} should be understood as a
criterion for when the intrinsic quantum fluctuations of the detector
mode are comparable to or larger than the gravitational-wave-induced
motion, rather than as a universal requirement for quantum sensing. In
this regime, the gravitational-wave perturbation cannot be regarded
solely as a classical displacement of the detector trajectory; instead,
the intrinsic quantum dynamics of the detector remains relevant
throughout the interaction.

As an illustrative example, consider a GHz-frequency mesoscopic
mechanical resonator such as a high-overtone bulk acoustic resonator
(HBAR)~\cite{Chu:2017koi} or an optomechanical crystal
(OMC)~\cite{OConnell2010,Teufel2011,Chan2011}. For a representative HBAR
with
$\omega/2\pi\simeq5.96~\mathrm{GHz}$
and effective mass
$m_{\rm eff}\simeq16.2~\mu\mathrm{g}$,
the corresponding zero-point displacement is
$x_{\rm zpf}\simeq2.9\times10^{-19}\,\mathrm{m}$
\cite{PhysRevD.111.026009}. For lighter optomechanical resonators,
zero-point displacements can reach values of order
$x_{\rm zpf}\sim10^{-17}\,\mathrm{m}$, further enhancing their quantum
response.

For comparison, a gravitational wave with representative strain
$h\sim10^{-23}$ acting over a micron-scale device,
$L\lesssim1\,\mu\mathrm{m}$, produces a displacement of order
\begin{equation}
\Delta x_{\rm GW}
=
hL
\sim
10^{-29}\,\mathrm{m}.
\end{equation}
For the representative parameters above, this gives
\begin{equation}
\frac{\Delta x_{\rm GW}}
{x_{\rm zpf}}
\sim
10^{-11}.
\end{equation}
Thus, for such a microscopic detector size, the gravitational-wave-induced
classical displacement is exceedingly small compared with the detector's
zero-point motion. Rather than classicalizing the detector dynamics or
significantly populating mechanical excitations, the gravitational
perturbation acts as a weak quantum interaction on an otherwise coherent
detector system.

This observation changes the relevant experimental question. For a
conventional interferometer, the primary observable is the classical
displacement or phase shift induced by the gravitational wave. For a
coherent mesoscopic quantum detector, the more relevant observables are
the quantum correlations generated during the detector--field interaction.
In particular, one may investigate quantities such as coherence,
detector--detector entanglement, von Neumann entropy, purity, and quantum
memory. These observables probe the quantum structure of the joint
detector--gravitational-field dynamics rather than merely the magnitude of
the classical tidal displacement.

This distinction can be summarized schematically as
\begin{equation}
\begin{array}{ccc}
\text{conventional interferometer}
&
\longrightarrow
&
\text{measure classical GW displacement},
\\[4pt]
\text{quantum detector}
&
\longrightarrow
&
\text{probe quantum correlations with the GW field}.
\end{array}
\end{equation}

If the gravitational-wave field is treated as classical, its role is that
of an externally prescribed tidal field, possibly supplemented by
classical stochastic fluctuations. Such fluctuations can modify the
statistics and purity of a quantum detector, but they do not by themselves
provide a quantum subsystem capable of mediating entanglement between
independently quantized detectors within the independent-mediator framework
considered here. By contrast, if the gravitational-wave field is
quantized, it becomes a dynamical quantum degree of freedom that can
participate in the joint unitary evolution and mediate quantum correlations
between detector subsystems.

The central question therefore changes from
``How accurately can a gravitational wave be measured?''
to
``Can a propagating gravitational-wave field coherently exchange quantum
information with quantum systems?''

Addressing this question requires treating both the detector degrees of
freedom and the gravitational-wave field within a common quantum-mechanical
framework. In particular, the detector must be described as a genuine
quantum subsystem rather than merely as a classical test mass. This allows
one to compare, within the same geometric interaction, the predictions of
a classical stochastic gravitational-wave field with those of a quantized
propagating field.

Motivated by these considerations, the following sections develop a
unified theoretical framework for localized quantum detectors interacting
with a propagating gravitational-wave field. Starting from the geodesic
deviation equation in linearized General Relativity, we derive the
effective detector Hamiltonian directly from spacetime geometry rather
than introducing a phenomenological interaction. The resulting framework
naturally describes two independent transverse detector degrees of
freedom, which are subsequently quantized as two harmonic-oscillator
subsystems. We then compare their dynamics under classical and quantized
gravitational-wave fields and investigate the resulting quantum
correlations, including detector--detector entanglement, coherence,
entropy, purity, and quantum memory.

%%%%%%%%%%%%%%%%%%%%%%%%%%%%%%%%%%%%%%%%%%%%%%%%%%%%%%%%%%%%%%%
\section{Information Transfer Between Quantum Systems and Physical Fields}
\label{sec:information_transfer}

The previous section established the motivation for employing quantum
detectors to probe possible quantum properties of propagating
gravitational waves. An equally important question concerns how
quantum information is exchanged between a quantum system and the
physical field with which it interacts. Before constructing the
gravitational-wave detector model, we therefore briefly review the
general operational paradigm through which physical fields can
acquire, transmit, and store information about quantum systems.

One of the central principles of quantum measurement is that
information about a microscopic quantum system is obtained through
its interaction with another physical degree of freedom. A
well-controlled probe is prepared, allowed to interact with the
target system, and subsequently measured. During the interaction,
the probe and the target generally become correlated, so that
information about the state of the target is encoded in the probe.
This probe-based perspective underlies a wide range of quantum
measurements and scattering experiments.

\begin{figure*}[t]
    \centering
    \includegraphics[width=0.95\textwidth]
    {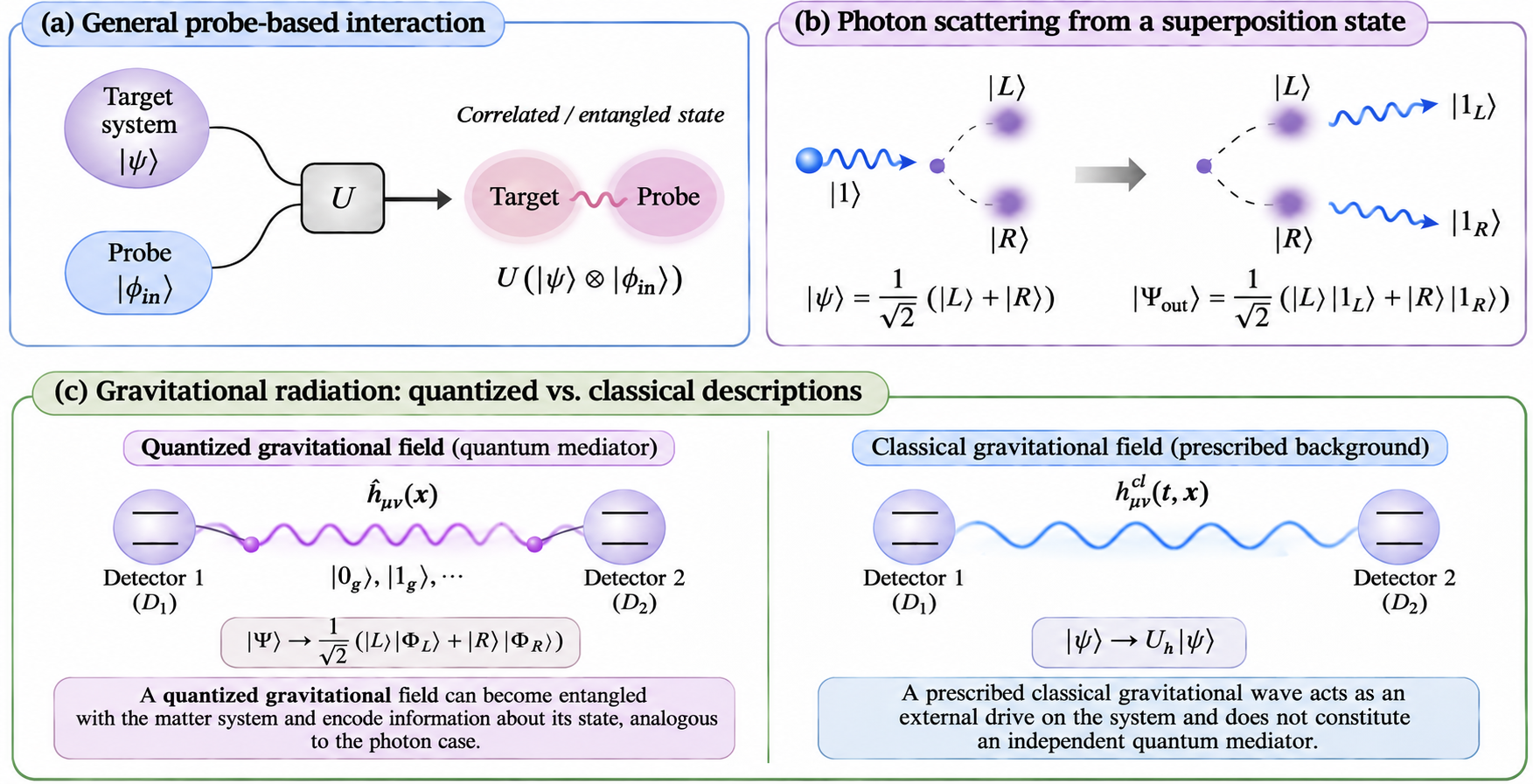}
    \caption{
    Schematic illustration of information transfer between
    quantum systems and physical fields.
    (a) General probe-based interaction, in which a quantum
    target and a physical probe become correlated through
    their joint evolution.
    (b) Photon scattering from a quantum particle prepared
    in a superposition of two distinguishable spatial
    branches, illustrating the correlation between the
    matter branches and the corresponding outgoing photon
    states.
    (c) Gravitational-field interaction with quantum matter.
    In the quantized description (left), the gravitational
    field can become entangled with the matter system, with
    different field states correlated with different matter
    branches. In the classical description (right), the
    gravitational wave is represented by a prescribed
    classical background rather than an independent quantum
    degree of freedom. This comparison motivates the
    conceptual questions examined in the Eppley--Hannah
    thought experiment.
    }
    \label{fig:information_transfer}
\end{figure*}

The general structure of such an interaction can be expressed
through the joint unitary evolution
\begin{equation}
|\psi\rangle\otimes|\phi_{\rm in}\rangle
\longrightarrow
U\left(
|\psi\rangle\otimes|\phi_{\rm in}\rangle
\right),
\end{equation}
where $|\psi\rangle$ denotes the state of the target system,
$|\phi_{\rm in}\rangle$ is the initial state of the probe, and
$U$ denotes the joint evolution generated by their interaction.
In general, the final state need not be separable: the probe
becomes correlated with the target and can consequently carry
information about its quantum state. This is the basic mechanism
underlying the transfer of quantum information through a physical
probe, as schematically illustrated in
Fig.~\ref{fig:information_transfer}(a).

A familiar example is provided by photon scattering from a
quantum system. Consider, for illustration, a particle prepared
in the superposition
\begin{equation}
|\psi\rangle
=
\frac{1}{\sqrt{2}}
\left(
|L\rangle+|R\rangle
\right),
\end{equation}
where $|L\rangle$ and $|R\rangle$ denote two distinguishable
spatial branches of the same quantum particle. If a single photon
is used as the probe, the initial state of the combined system is
\begin{equation}
|\Psi_{\rm in}\rangle
=
|\psi\rangle\otimes|1\rangle.
\end{equation}
Following the interaction, the joint state can schematically
take the form
\begin{equation}
|\Psi_{\rm out}\rangle
=
\frac{1}{\sqrt{2}}
\left(
|L\rangle\otimes|1_L\rangle
+
|R\rangle\otimes|1_R\rangle
\right),
\end{equation}
where $|1_L\rangle$ and $|1_R\rangle$ denote the corresponding
outgoing photon states. The photon therefore becomes correlated
with the different branches of the particle's quantum
superposition and can carry information about its state.
The essential feature is that the probe itself is a quantum
degree of freedom and can consequently become entangled with
the matter system during the interaction. This process is
illustrated in Fig.~\ref{fig:information_transfer}(b).

The same general question can be raised for gravitational
radiation. If the gravitational field is quantized, its
interaction with a quantum matter system can be described
in an analogous manner. For a matter system initially prepared
in the superposition
\begin{equation}
|\psi\rangle
=
\frac{1}{\sqrt{2}}
\left(
|L\rangle+|R\rangle
\right),
\end{equation}
and a gravitational field initially in a state
$|\Phi_g\rangle$, the joint evolution may schematically
produce
\begin{equation}
|\Psi_{\rm out}\rangle
=
\frac{1}{\sqrt{2}}
\left(
|L\rangle\otimes|\Phi_L\rangle
+
|R\rangle\otimes|\Phi_R\rangle
\right),
\end{equation}
where $|\Phi_L\rangle$ and $|\Phi_R\rangle$ are the
gravitational-field states correlated with the two matter
branches. In this description, the gravitational field can
become entangled with the matter system and thereby carry
information about its quantum state, in direct analogy with
the photon-scattering example.

The distinction between quantized and classical gravitational
radiation is illustrated schematically in
Fig.~\ref{fig:information_transfer}(c). In the quantized
description, the gravitational field possesses its own
quantum degrees of freedom and can become correlated with
the matter system. By contrast, a prescribed classical
gravitational background is represented by a classical
field configuration, rather than by an independent quantum
state. Consequently, within this prescribed-background
description, there are no gravitational-field quantum states
that become entangled with the matter branches. The question
of how a fundamentally classical gravitational field can
consistently interact with quantum matter, while retaining
the essential structure of quantum mechanics, is therefore
a nontrivial one.

This issue was famously explored by Eppley and
Hannah~\cite{Eppley:1977emg}, who considered the interaction
of a classical gravitational wave with a quantum particle.
Their thought experiment was intended to investigate whether
a fundamentally classical gravitational field could
consistently serve as a probe of a quantum system. They
argued that a conceptual dilemma arises: if the gravitational
field acquires information about the quantum state, the
interaction appears to act as a measurement and can lead to
wavefunction collapse; if, instead, the quantum superposition
is preserved, it becomes unclear how a classical field can
retain information about the different branches of the
superposition without possessing quantum degrees of freedom.

The Eppley--Hannah argument, however, is not generally
regarded as a conclusive proof that gravity must be
quantized. Subsequent analyses have emphasized its
reliance on idealized assumptions concerning classical
gravitational probes, measurement precision, backreaction,
and the formulation of a consistent hybrid classical--quantum
dynamics. Consequently, the thought experiment remains
conceptually influential but does not by itself provide
a definitive operational criterion for establishing the
quantum nature of gravity.

The present work adopts a complementary operational
perspective. Rather than attempting to determine the quantum
state of matter by measuring an outgoing gravitational wave,
we investigate whether a propagating gravitational-wave
field can act as a common mediator of quantum correlations
between spatially separated quantum detector subsystems.
This is a different physical arrangement from the
single-particle probe scenario discussed above. Here, the
gravitational field interacts locally with two distinct
quantum systems, and the relevant question is whether their
mutual interaction with the common field can generate
detector--detector entanglement.

In the quantized description, the two detectors interact
with a common gravitational degree of freedom and can
therefore develop correlations through their interactions
with the field. The gravitational field is treated as an
intermediate physical degree of freedom, while the
experimentally relevant observables are properties of
the detector subsystems after the interaction. Whether
these correlations include genuine detector--detector
entanglement is the question addressed by the dynamical
analysis below. By contrast, a prescribed classical
gravitational background acts as an external classical
drive and does not constitute an independent quantum
mediator in the description considered here.

This operational viewpoint provides the conceptual bridge
to the detector model developed in the next section.
We first derive the detector--gravitational-wave interaction
directly from the geometry of General Relativity, using
geodesic deviation to characterize the tidal response of
localized systems. The resulting effective Hamiltonian
then provides a common starting point for the classical
and quantized descriptions of gravitational-wave-mediated
dynamics.
%%%%%%%%%%%%%%%%%%%%%%%%%%%%%%%%%%%%%%%%%%%%%%%%%%%%%%%%%%%%%%%

\section{Geometry of Gravitational Waves and Quantum Detector Dynamics}
%%%%%%%%%%%%%%%%%%%%%%%%%%%%%%%%%%%%%%%%%%%%%%%%%%%%%%%%%%%%%%%

Having established the general paradigm through which physical fields can
exchange information with quantum systems, we now construct the corresponding
detector model for gravitational waves. Unlike electromagnetic interactions,
which are described in terms of interaction potentials, gravity is encoded in
the geometry of spacetime itself. Consequently, the coupling between a quantum
detector and a gravitational wave must emerge from the geometric description
of particle motion in curved spacetime rather than being introduced
phenomenologically. In this section, we derive the effective detector
Hamiltonian from General Relativity, thereby establishing the geometric
foundation for the classical and quantum gravitational-wave dynamics
considered in the following sections.

The natural starting point is the motion of a freely falling test particle.
In General Relativity, the trajectory of such a particle is obtained by
extremizing the proper-time action,
\begin{equation}
S_{\rm p}
=
-mc
\int ds
=
-mc
\int
\sqrt{
g_{\mu\nu}(x)
dx^\mu dx^\nu},
\label{action}
\end{equation}
where $g_{\mu\nu}$ is the spacetime metric. Variation of this action with
respect to the particle trajectory yields the geodesic equation,
\begin{equation}
\frac{d^2x^\mu}{d\tau^2}
+
\Gamma^\mu_{\nu\rho}
\frac{dx^\nu}{d\tau}
\frac{dx^\rho}{d\tau}
=
0,
\label{geodesic}
\end{equation}
where
\begin{equation}
\Gamma^\mu_{\nu\rho}
=
\frac12
g^{\mu\lambda}
\left(
\partial_\nu g_{\lambda\rho}
+
\partial_\rho g_{\lambda\nu}
-
\partial_\lambda g_{\nu\rho}
\right)
\end{equation}
are the Christoffel symbols associated with the spacetime metric.

Equation~(\ref{geodesic}) describes the motion of a single freely falling
particle. However, the local effects of gravity can be removed along an
individual freely falling worldline by an appropriate choice of locally
inertial coordinates. Gravitational effects that cannot be removed in this
way are therefore encoded in spacetime curvature and manifest themselves
through the relative motion of neighbouring freely falling particles. This
tidal response is described by the geodesic deviation equation and provides
the appropriate starting point for modelling the response of a localized
gravitational-wave detector.

To characterize this relative motion, consider two neighbouring geodesics
$x^\mu(\tau)$ and $x^\mu(\tau)+q^\mu(\tau)$, where $q^\mu$ denotes the
infinitesimal separation vector connecting the two worldlines. The evolution
of this separation is governed by
\begin{equation}
\frac{D^2q^\mu}{D\tau^2}
=
-
R^\mu_{\;\nu\rho\sigma}
u^\nu
q^\rho
u^\sigma,
\label{GD}
\end{equation}
where
$u^\mu=dx^\mu/d\tau$
is the four-velocity of the reference geodesic and
$R^\mu_{\;\nu\rho\sigma}$
is the Riemann curvature tensor. Unlike the geodesic equation, which
describes the trajectory of a single particle, Eq.~(\ref{GD}) characterizes
the relative acceleration generated by spacetime curvature and therefore
captures the tidal effect relevant for gravitational-wave detection.

We consider weak gravitational waves propagating on an approximately flat
background. The metric is written as
\begin{equation}
g_{\mu\nu}
=
\eta_{\mu\nu}
+
h_{\mu\nu},
\qquad
|h_{\mu\nu}|\ll1,
\label{metric}
\end{equation}
where $\eta_{\mu\nu}$ is the Minkowski metric and $h_{\mu\nu}$ denotes the
metric perturbation. Throughout this work, we retain terms only to first
order in the gravitational-wave perturbation.

Choosing the transverse-traceless (TT) gauge,
\begin{equation}
h_{0\mu}=0,
\qquad
\partial^j h_{ij}=0,
\qquad
h^i_{\ i}=0,
\end{equation}
isolates the two physical polarization degrees of freedom of the
gravitational wave. For a wave propagating along the $z$ direction and for
slowly moving test particles,
$u^\mu\simeq(c,0,0,0)$
and
$d\tau\simeq dt$,
the spatial components of Eq.~(\ref{GD}) reduce to
\begin{equation}
\frac{d^2q^i}{dt^2}
=
-c^2
R^i_{\;0j0}
q^j.
\label{NRGD}
\end{equation}
Using the linearized Riemann tensor in the TT gauge,
\begin{equation}
R_{0i0j}
=
-\frac12
\ddot h^{\rm TT}_{ij},
\end{equation}
we obtain
\begin{equation}
\ddot q^i(t)
=
\frac12
\ddot h^{\rm TT}_{ij}(t)
q^j(t).
\label{tidal}
\end{equation}
Equation~(\ref{tidal}) describes the tidal acceleration induced by the
gravitational wave and provides the geometric starting point for the
detector model developed below.

To turn this freely falling relative coordinate into a localized quantum
detector, we introduce an external trapping potential $V(\mathbf q)$ that
confines the system around a stable equilibrium position. The trap provides
the restoring force that defines the detector's internal quantum degree of
freedom, while the gravitational wave acts as a weak time-dependent tidal
perturbation. The resulting equation of motion is
\begin{equation}
m\ddot q_i
=
\frac{m}{2}
\ddot h^{\rm TT}_{ij}(t)
q_j
+
F_i^{\rm ex},
\qquad i=1,2,
\label{detectorEOM}
\end{equation}
where
\begin{equation}
F_i^{\rm ex}
=
-
\frac{\partial V}{\partial q_i}.
\end{equation}
This provides an effective description applicable to a broad class of
localized quantum systems, including trapped mechanical oscillators,
optomechanical resonators, levitated systems, and related quantum sensing
platforms.

%%%%%%%%%%%%%%%%%%%%%%%%%%%%%%%%%%%%%%%%%%%%%%%%%%%%%%%%%%%%%%%
\subsection{Effective Detector Hamiltonian}
%%%%%%%%%%%%%%%%%%%%%%%%%%%%%%%%%%%%%%%%%%%%%%%%%%%%%%%%%%%%%%%

We now construct the Hamiltonian corresponding to the detector dynamics in
Eq.~(\ref{detectorEOM}). For definiteness, we consider a harmonic trapping
potential,
\begin{equation}
V(\mathbf q)
=
\frac12m\Omega_0^2q_iq_i,
\end{equation}
where $\Omega_0$ is the natural frequency of the detector. The harmonic
trap provides a simple and experimentally relevant realization of the
localized quantum detector introduced above. With this potential, the
equation of motion~(\ref{detectorEOM}) contains both the restoring force of
the trap and the gravitational-wave-induced tidal force.

To first order in the gravitational-wave perturbation, the corresponding
Lagrangian can be written as
\begin{equation}
L
=
\frac12
m\dot q_i\dot q_i
-
\frac12
m\Omega_0^2q_iq_i
-
\frac{m}{2}
\dot h^{\rm TT}_{ij}(t)
q_i\dot q_j.
\label{Lag}
\end{equation}
The first term represents the kinetic energy of the detector, the second
describes the harmonic confinement, and the final term gives the
gravitational-wave interaction. Thus, the detector--gravity coupling is
derived from the tidal dynamics of spacetime rather than introduced as an
independent phenomenological interaction.

The canonical momentum conjugate to $q_i$ is
\begin{equation}
p_i
=
\frac{\partial L}{\partial\dot q_i}
=
m\dot q_i
-
\frac{m}{2}
\dot h^{\rm TT}_{ij}(t)
q_j.
\label{momentum}
\end{equation}
The Hamiltonian follows from the Legendre transformation,
\begin{equation}
H
=
p_i\dot q_i-L,
\end{equation}
which, to first order in the gravitational-wave amplitude, yields
\begin{equation}
H
=
\frac{p_i^2}{2m}
+
\frac12
m\Omega_0^2q_iq_i
+
\frac12
\dot h^{\rm TT}_{ij}(t)
q_ip_j.
\label{DetectorHamiltonian}
\end{equation}
The first two terms define the free Hamiltonian of the trapped detector,
while the final term,
\begin{equation}
H_{\rm int}
=
\frac12
\dot h^{\rm TT}_{ij}(t)q_ip_j,
\label{geometricinteraction}
\end{equation}
describes the interaction between the detector and the gravitational wave.

Equation~(\ref{DetectorHamiltonian}) provides the common detector--field
Hamiltonian used in the subsequent analysis. If $h^{\rm TT}_{ij}(t)$ is
treated as a prescribed classical function, Eq.~(\ref{DetectorHamiltonian})
describes a quantum detector driven by an external classical gravitational
field. If instead the gravitational-wave perturbation is promoted to a
quantum field operator, the same geometric coupling becomes an interaction
between the detector and a dynamical gravitational degree of freedom. Thus,
the classical and quantized descriptions share the same underlying
detector--gravity coupling; they differ only in the physical description of
the gravitational field.

For the plus polarization considered here, the two transverse detector
degrees of freedom can be represented by two independent oscillator modes.
We therefore identify
\begin{equation}
\mathcal H_D
=
\mathcal H_1\otimes\mathcal H_2,
\end{equation}
where $\mathcal H_1$ and $\mathcal H_2$ denote the Hilbert spaces of the two
detector modes. The absence of a direct interaction between these modes is
important for the interpretation of the correlations discussed later.

Indeed, Eq.~(\ref{geometricinteraction}) contains no direct
detector--detector coupling. Each detector couples locally to the
gravitational field, while the gravitational field provides the common
degree of freedom shared by the two detector subsystems. For a prescribed
classical gravitational-wave background, the detector evolution therefore
acts locally on the two detector modes and factorizes as
\begin{equation}
\hat U_{\rm int}^{\gamma}(t)
=
\hat U_{\rm int,1}^{\gamma}(t)
\otimes
\hat U_{\rm int,2}^{\gamma}(t).
\label{UU}
\end{equation}
Consequently, an initially separable detector state remains separable under
such local classical driving. This provides the reference point against
which the quantized gravitational-field description will be compared.

The distinction becomes important when the gravitational field itself is
treated as a quantum degree of freedom. In that case, both detector modes
couple to the same gravitational subsystem. The gravitational field can
therefore act as a common quantum mediator and can become correlated with
both detectors. Tracing out the gravitational degrees of freedom can then
produce a nontrivial reduced state of the two detectors. Establishing
whether this reduced state contains genuine detector--detector entanglement
is the central objective of the subsequent analysis.

The Hamiltonian in Eq.~(\ref{DetectorHamiltonian}) therefore provides the
geometric starting point for both descriptions considered in this work. In
the next section, we quantize the gravitational field and formulate the
corresponding detector--graviton dynamics. We first establish the
interaction-picture Hamiltonian and the effective single-mode description,
and subsequently use a perturbative Magnus expansion to determine the
joint detector--graviton state and its reduced detector dynamics.

\section{System Hamiltonian and quantum gravitational interaction}
\label{sec:H}

As discussed in the previous section, the interaction between a
gravitational wave and a localized system originates from the
geodesic deviation equation, which describes the tidal response of
spacetime. For a gravitational wave propagating along the $z$-axis,
the relevant response takes place in the transverse $(x,y)$ plane
and is described by the two components of the separation vector.
Accordingly, our starting point is a single localized system with
two transverse dynamical degrees of freedom. Upon introducing a
trapping potential, these degrees of freedom can be represented by
two effective harmonic oscillator modes, which we subsequently
quantize. We therefore introduce the effective bipartite Hilbert
space
\begin{equation}
\mathcal{H}_{D}
=
\mathcal{H}_{D_x}\otimes\mathcal{H}_{D_y},
\end{equation}
where $D_x$ and $D_y$ denote the two transverse oscillator modes
of the same localized system. The tensor-product structure defines
the mode bipartition across which we investigate the generation
of quantum correlations. The use of distinct modes of a single
physical system as quantum subsystems is well established in
quantum optics, where entanglement between different modes of
the same cavity has been experimentally demonstrated
\cite{PhysRevA.64.050301}. In particular, the two modes considered
here are distinct dynamical subsystems, but they are not
spatially separated detectors.

Within the long-wavelength approximation, the gravitational
wavelength is assumed to be much larger than the characteristic
size of the localized system. The gravitational-wave field can
therefore be treated as effectively uniform across the system.
Consequently, the relevant detector dynamics are described by
the two transverse oscillator degrees of freedom.

In the weak-field approximation, the dynamics of these two
oscillator degrees of freedom under a gravitational wave
propagating along the $z$-axis are governed by the time-dependent
Hamiltonian \cite{Nandi:2022sjy}
\begin{eqnarray}
 \hat{H}(t) &=&
 \sum_{i=1}^{2}
 \left(
 \alpha \hat{p}_i^2+\beta \hat{q}_i^2
 \right)
 +\gamma(t)
 \left(
 \hat{q}_1\hat{p}_1+\hat{p}_1\hat{q}_1
 \right)
 \nonumber\\
&&
-\gamma(t)
\left(
\hat{q}_2\hat{p}_2+\hat{p}_2\hat{q}_2
\right)
+\delta(t)
\left(
\hat{q}_1\hat{p}_2+\hat{p}_1\hat{q}_2
\right),
\label{eq:GW_Hamiltonian}
\end{eqnarray}
where
\begin{equation}
 \alpha=\frac{1}{2m},
 \qquad
 \beta=\frac{1}{2}m\Omega_0^2,
 \qquad
 \gamma(t)=\frac{\dot{\chi}(t)}{2}\epsilon_+,
 \qquad
 \delta(t)=\dot{\chi}(t)\epsilon_\times .
\end{equation}
In the transverse-traceless gauge, the gravitational-wave metric
perturbation is
\begin{equation}
 h_{jk}(t)
 =
 2\chi(t)
 \left(
 \epsilon_\times\sigma^1_{jk}
 +
 \epsilon_+\sigma^3_{jk}
 \right),
\end{equation}
where $\sigma^1$ and $\sigma^3$ are Pauli matrices, and
$\epsilon_+$ and $\epsilon_\times$ denote the plus and cross
polarization amplitudes, respectively. Here, $2\chi(t)$ denotes
the time-dependent gravitational-wave amplitude. A detailed
derivation of the Hamiltonian is given in
\cite{PhysRevD.51.1701,Nandi:2024jyf}; see also the appendices of
\cite{Dutta:2025ouy,Dutta:2025bge}.

The cross-polarization contribution, proportional to $\delta(t)$,
couples the two transverse oscillator modes directly. By contrast,
the plus-polarization contributions act separately on the two
modes. Thus, if the cross-polarization term is retained, the
Hamiltonian already contains a direct mode--mode interaction,
which can generate correlations even when the gravitational
wave is treated as a prescribed classical field. Such a
contribution would make it difficult to isolate correlations
mediated by the quantum gravitational field.

To separate these two mechanisms, we choose the transverse mode
axes to coincide with the principal axes of the gravitational-wave
tidal tensor. For the fixed polarization orientation considered
here, the transverse tidal tensor can be diagonalized by a
time-independent rotation in the $(x,y)$ plane. In this
principal-axis basis, the cross-polarization contribution
vanishes. This choice is a restriction to the fixed-polarization
sector of the model; it is not a statement that the cross
polarization is absent for a general gravitational wave with
an arbitrary or time-dependent polarization orientation.

We therefore work in this principal-axis basis and set
$\epsilon_\times=0$, so that $\delta(t)=0$. The resulting
Hamiltonian takes the form
\begin{equation}
\hat{H}(t)
=
\hat{H}_{D_x}(t)\otimes\mathbb{I}_{D_y}
+
\mathbb{I}_{D_x}\otimes\hat{H}_{D_y}(t),
\label{eq:H_principal_axis}
\end{equation}
where
\begin{align}
\hat{H}_{D_x}(t)
&=
\alpha\hat{p}_1^2+\beta\hat{q}_1^2
+\gamma(t)
\left(
\hat{q}_1\hat{p}_1+\hat{p}_1\hat{q}_1
\right),
\\
\hat{H}_{D_y}(t)
&=
\alpha\hat{p}_2^2+\beta\hat{q}_2^2
-\gamma(t)
\left(
\hat{q}_2\hat{p}_2+\hat{p}_2\hat{q}_2
\right).
\end{align}
Here, $\mathbb{I}_{D_x}$ and $\mathbb{I}_{D_y}$ are the identity
operators on the corresponding mode Hilbert spaces. Equation
(\ref{eq:H_principal_axis}) makes explicit that, in the chosen
basis, the prescribed gravitational-wave interaction contains
no direct coupling between the two transverse modes. The
gravitational-wave-induced terms act separately on the two
mode Hilbert spaces. In this sense, the interaction is
mode-local; this terminology refers to the mode decomposition
and does not imply that the two modes are spatially separated.

For a prescribed classical gravitational wave, the detector
Hamiltonian is therefore a sum of operators acting on the
individual mode Hilbert spaces. Since these operators act on
different tensor factors, the corresponding interaction-picture
evolution operator factorizes:
\begin{equation}
 \hat{U}^{\gamma}_{\mathrm{int}}(t)
 =
 \hat{U}^{\gamma}_{\mathrm{int},x}(t)
 \otimes
 \hat{U}^{\gamma}_{\mathrm{int},y}(t).
\label{UU}
\end{equation}
Consequently, an initially separable state of the two transverse
modes remains separable under the evolution generated by a
prescribed classical gravitational wave. The same conclusion
holds for a classical stochastic gravitational-wave background
described as an ensemble of such prescribed-field evolutions:
although averaging over different realizations can produce a
mixed detector state, it cannot generate entanglement between
the two modes from an initially separable detector state.

The situation is different when the gravitational field is
treated as a quantum dynamical degree of freedom. In this case,
the two transverse modes interact with the same quantized
gravitational-field degrees of freedom. The field thus provides
a common quantum mediator through which correlations can be
generated between the modes, even though the direct
mode--mode interaction has been removed from the detector
Hamiltonian in the principal-axis basis. In the following
sections, we investigate the resulting reduced dynamics of
the two transverse modes and determine whether the
quantized-field-mediated interaction generates entanglement
across the bipartition
$\mathcal{H}_{D_x}\otimes\mathcal{H}_{D_y}$.

Our analysis therefore compares two evolutions within the
specified interaction model: one in which the gravitational
wave is a prescribed classical field and one in which the
gravitational field is quantized and acts as a common
dynamical mediator. The resulting mode entanglement is
interpreted as a signature of the quantized-field-mediated
dynamics under these assumptions, rather than as a
model-independent proof of the quantization of gravity.

\subsection{Quantized gravitational interaction}

We now quantize the gravitational-wave degree of freedom. The
annihilation operators for the detector oscillators are defined by
\begin{equation}
 \hat{a}_i
 =
 \left(\frac{\alpha}{\beta}\right)^{1/4}
 \frac{
 \sqrt{\beta/\alpha}\,\hat{q}_i+i\hat{p}_i
 }{\sqrt{2\hbar}},
 \qquad i=1,2,
\end{equation}
with
\begin{equation}
 [\hat{a}_i,\hat{a}_j^\dagger]
 =
 \delta_{ij}\mathbb{I}.
\end{equation}

The quantized gravitational field is described by bosonic mode
operators $\hat b_k$ and $\hat b_k^\dagger$, satisfying
\begin{equation}
 [\hat b_k,\hat b_{k'}^\dagger]
 =
 \delta_{kk'}.
\end{equation}
The free Hamiltonian of the gravitational field is
\begin{equation}
 \hat H_G
 =
 \sum_k
 \hbar\omega_k
 \left(
 \hat b_k^\dagger\hat b_k+\frac{1}{2}
 \right).
\end{equation}

In the effective single-mode description adopted below, we retain one
mode, denoted by $g$, with frequency $\omega_g$. The corresponding
free Hamiltonian is therefore
\begin{equation}
 \hat H_G
 =
 \hbar\omega_g
 \left(
 \hat b_g^\dagger\hat b_g+\frac{1}{2}
 \right).
\end{equation}
A detailed derivation of the free gravitational Hamiltonian is given
in Appendix~\ref{Appendix A}.

At this stage all operators are understood as Schrödinger-picture
operators. In particular, the gravitational interaction operator
associated with the retained mode is
\begin{equation}
 \hat{\gamma}_S
 =
 iC_\gamma
 \left(
 \hat b_g-\hat b_g^\dagger
 \right),
\label{eq:gamma_S}
\end{equation}
where
\begin{equation}
 C_\gamma
 =
 -\sqrt{
 \frac{\omega_g c\pi l_p^2}{2L^3}
 },
\end{equation}
with $L$ denoting the quantization-box size and $l_p$ the Planck
length.

The total Hamiltonian of the detector--graviton system in the
Schrödinger picture is
\begin{equation}
 \hat H_S
 =
 \mathbb I_D\otimes\hat H_G
 +
 \hat H_D\otimes\mathbb I_G
 +
 \hat H_{\mathrm{int},S},
\label{eq:total_Hamiltonian_S}
\end{equation}
where
\begin{equation}
 \hat H_D
 =
 2\hbar\sqrt{\alpha\beta}
 \left(
 \sum_{i=1}^{2}\hat N_i+1
 \right),
 \qquad
 \hat N_i=\hat a_i^\dagger\hat a_i,
\end{equation}
and
\begin{equation}
 \hat H_{\mathrm{int},S}
 =
 i\hbar
 \left(
 \hat a_1^{\dagger 2}
 -\hat a_1^2
 -\hat a_2^{\dagger 2}
 +\hat a_2^2
 \right)
 \otimes\hat\gamma_S .
\label{eq:Hint_S}
\end{equation}

%\subsection{Interaction-picture description}

We now pass to the interaction picture with respect to the complete
free Hamiltonian of the detector--graviton system,
\begin{equation}
 \hat H_0
 =
 \mathbb I_D\otimes\hat H_G
 +
 \hat H_D\otimes\mathbb I_G .
\label{eq:H0}
\end{equation}
The interaction-picture Hamiltonian is defined by
\begin{equation}
 \hat H_{\mathrm{int}}^I(t)
 =
 e^{\frac{i}{\hbar}\hat H_0t}
 \hat H_{\mathrm{int},S}
 e^{-\frac{i}{\hbar}\hat H_0t}.
\label{eq:HI_definition}
\end{equation}

Because $\hat H_D$ and $\hat H_G$ act on different Hilbert spaces,
their free evolutions commute. The detector operators consequently
evolve according to
\begin{equation}
 \hat a_i^I(t)
 =
 e^{\frac{i}{\hbar}\hat H_Dt}
 \hat a_i
 e^{-\frac{i}{\hbar}\hat H_Dt}
 =
 \hat a_i e^{-i\Omega_0t},
\end{equation}
where
\begin{equation}
 \Omega_0=2\sqrt{\alpha\beta}.
\end{equation}
Hence,
\begin{equation}
 \hat a_i^{I\,2}(t)
 =
 \hat a_i^2e^{-2i\Omega_0t},
 \qquad
 \hat a_i^{I\,\dagger 2}(t)
 =
 \hat a_i^{\dagger 2}e^{2i\Omega_0t}.
\end{equation}

Similarly, the gravitational mode evolves under $\hat H_G$ as
\begin{equation}
 \hat b_g^I(t)
 =
 e^{\frac{i}{\hbar}\hat H_Gt}
 \hat b_g
 e^{-\frac{i}{\hbar}\hat H_Gt}
 =
 \hat b_g e^{-i\omega_gt},
\end{equation}
and
\begin{equation}
 \hat b_g^{I\dagger}(t)
 =
 \hat b_g^\dagger e^{i\omega_gt}.
\end{equation}
Therefore, the interaction-picture gravitational operator is
\begin{equation}
 \hat\gamma_I(t)
 =
 iC_\gamma
 \left(
 \hat b_g e^{-i\omega_gt}
 -
 \hat b_g^\dagger e^{i\omega_gt}
 \right).
\label{eq:gamma_I}
\end{equation}
The factors $e^{\mp i\omega_gt}$ in Eq.~(\ref{eq:gamma_I}) arise from
the free gravitational evolution and are therefore included only once.

Combining the detector and gravitational free evolutions gives
\begin{equation}
\begin{aligned}
 \hat H_{\mathrm{int}}^I(t)
 ={}&
 i\hbar
 \Big[
 \hat a_1^{\dagger2}e^{2i\Omega_0t}
 -\hat a_1^2e^{-2i\Omega_0t}
\\
&\qquad
 -\hat a_2^{\dagger2}e^{2i\Omega_0t}
 +\hat a_2^2e^{-2i\Omega_0t}
 \Big]
 \otimes\hat\gamma_I(t).
\end{aligned}
\label{eq:HI_final}
\end{equation}

Equation~(\ref{eq:HI_final}) makes explicit the origin of all
time-dependent phases in the interaction picture. The detector
quadratic operators acquire phases determined by the detector
frequency, while the gravitational mode operators acquire phases
determined by the GW frequency. No additional interaction-picture
transformation of $\hat\gamma_I(t)$ is performed after
Eq.~(\ref{eq:gamma_I}).

%%%%%%%%%%%%%%%%%%%%%%%%%%%%%%%%%%%%%%%%%%%%%%%%%%%%%%%%%%%%%%%

\subsection{Single-mode approximation and multimode contributions}
%%%%%%%%%%%%%%%%%%%%%%%%%%%%%%%%%%%%%%%%%%%%%%%%%%%%%%%%%%%%%%%

It is useful to clarify the scope and physical meaning of the
single-mode approximation adopted in this work. The complete
quantized gravitational field contains a continuum of modes, and
the corresponding interaction Hamiltonian may be written
schematically as
\begin{equation}
\hat H_{\mathrm{int},S}
=
\sum_k \hat H_{\mathrm{int},S;k},
\label{MultimodeHint}
\end{equation}
where $k$ labels the gravitational-field modes. In the present
effective description, we retain a single gravitational mode,
denoted by $g$, with frequency $\omega_g$. This provides a
simplified description in which one gravitational degree of
freedom is treated explicitly, allowing us to isolate the
detector--field dynamics relevant to the present calculation.
The remaining gravitational modes are not assumed to be
physically absent; rather, their contributions are not explicitly
included in the effective single-mode model. Their quantitative
effects, including possible modifications of the transition
amplitudes and dispersive corrections, require a separate
multimode analysis.

It is also important to distinguish the single-mode approximation
from the identification of resonant and off-resonant processes.
Consider first a detector transition in which the first oscillator
mode gains two excitation quanta,
\begin{equation}
|i\rangle
=
|n_1,n_2;N_k\rangle,
\qquad
|f\rangle
=
|n_1+2,n_2;N_k-1\rangle,
\end{equation}
where the initial gravitational mode is populated with $N_k$
gravitons. The detector transition frequency associated with
$n_1\rightarrow n_1+2$ is
\begin{equation}
\omega_0=2\Omega_0.
\end{equation}
For this graviton-absorption process, the energy difference is
\begin{equation}
E_f-E_i
=
\hbar(\omega_0-\omega_k),
\label{EnergyMismatch}
\end{equation}
so that the energy-conserving resonance condition is
\begin{equation}
\omega_k=\omega_0.
\label{OnShellCondition}
\end{equation}
Thus, real graviton absorption by the detector is enhanced in the
near-resonant region of the gravitational spectrum.

The same distinction is apparent in the interaction picture.
For an initially populated gravitational mode, the absorption
contribution contains the time-dependent factor
\begin{equation}
e^{2i\Omega_0t}e^{-i\omega_gt}
=
e^{i(\omega_0-\omega_g)t},
\label{AbsorptionFactor}
\end{equation}
and is therefore resonantly enhanced when
\begin{equation}
\omega_g\simeq\omega_0.
\end{equation}
Such processes describe the transfer of energy from an occupied
gravitational mode to the detector. They are also relevant to
proposed quantum-acoustic schemes for probing discrete energy
exchange between gravitational waves and quantum resonators
\cite{Tobar:2023ksi}.

The situation is different for the vacuum initial state considered
in the present calculation. Since the gravitational field initially
contains no gravitons, the relevant excitation process involves
the creation operator $\hat b_g^\dagger$. The corresponding
transition is
\begin{equation}
|n_1,n_2;0_g\rangle
\longrightarrow
|n_1+2,n_2;1_g\rangle.
\end{equation}
Its time dependence is
\begin{equation}
e^{2i\Omega_0t}e^{i\omega_gt}
=
e^{i(\omega_0+\omega_g)t}.
\label{VacuumCreationFactor}
\end{equation}
For positive $\omega_0$ and $\omega_g$, this contribution is
counter-rotating and has no positive-frequency on-shell
resonance. Nevertheless, it is retained in the present analysis
because the leading detector--detector correlations investigated
below originate from these vacuum-induced transition amplitudes.
Consequently, the present calculation does not employ a
rotating-wave approximation, even though the gravitational field
is restricted to one effective mode.

It is important to emphasize that the present analysis concerns
the vacuum-induced response of the detector modes to a quantized
propagating gravitational field. In particular, the initial
gravitational state is taken to be the vacuum, corresponding to
an initial occupation number
\begin{equation}
N_g
=
\langle 0_g|\hat b_g^\dagger\hat b_g|0_g\rangle
=
0.
\end{equation}
The detector excitation pathways discussed above therefore
originate from vacuum fluctuations of the gravitational mode
rather than from stimulated absorption or emission associated
with a highly populated gravitational-wave state.

For an initially populated gravitational mode, additional
contributions arise from the annihilation operator $\hat b_g$,
including near-resonant processes satisfying
$\omega_g\simeq 2\Omega_0$. These processes can modify the
detector excitation probabilities and the resulting reduced-state
coherences. However, their presence does not, by itself, establish
a quantum signature, since stimulated responses can also arise
in the corresponding classical-wave description. A systematic
comparison therefore requires treating the quantum and classical
descriptions for the same specified field state and distinguishing
the resulting detector correlations from genuine entanglement.

The vacuum configuration considered here thus isolates a
distinct physical question: whether vacuum fluctuations of a
propagating quantum gravitational degree of freedom can generate
quantum correlations between initially independent transverse
detector modes.

In a multimode treatment, the distinction between on-shell and
off-shell contributions can be expressed using the standard
distributional identity
\begin{equation}
\frac{1}{E_i-E_n+i\epsilon}
=
\mathcal{P}\frac{1}{E_i-E_n}
-
i\pi\delta(E_i-E_n),
\label{WWdecomposition}
\end{equation}
where $\mathcal{P}$ denotes the principal-value prescription.
This decomposition arises naturally in continuum treatments
of transition amplitudes, particularly in the long-time
Weisskopf--Wigner description. The delta-function contribution
selects energy-conserving transitions, whereas the principal-value
contribution describes off-resonant dispersive effects.

For the graviton-absorption process considered above, the
energy-conserving contribution is selected by
\begin{equation}
\delta(E_i-E_f)
=
\frac{1}{\hbar}
\delta(\omega_k-\omega_0).
\label{OnShellDelta}
\end{equation}
This makes explicit why real energy exchange is concentrated
near $\omega_k\simeq\omega_0$. By contrast, for the vacuum-induced
creation process,
\begin{equation}
E_f-E_i=\hbar(\omega_0+\omega_k),
\end{equation}
and the corresponding energy-conserving delta function has no
support for positive detector and gravitational frequencies.
The vacuum-induced contribution is therefore off shell in the
energy-conserving sense.

Off-resonant modes can nevertheless contribute to the finite-time
response. In the continuum description, their dispersive
contributions are associated with principal-value terms. For
example, if the coupling to mode $k$ is characterized by $g_k$,
a typical second-order energy correction has the schematic form
\begin{equation}
\Delta E_i^{\mathrm{virt}}
\sim
\mathcal{P}
\sum_k
\frac{
\left|
\langle n|
\hat H_{\mathrm{int},S;k}
|i\rangle
\right|^2
}
{E_i-E_n}
\sim
\mathcal{P}
\sum_k
\frac{|g_k|^2}{E_i-E_n},
\label{VirtualContribution}
\end{equation}
up to the appropriate detector matrix elements and numerical
factors. Such terms can contribute to dispersive corrections,
including shifts of detector energies or effective transition
frequencies. At finite times, off-resonant processes may also
contribute to transient transition amplitudes and coherences.
They should therefore not be identified with additional
energy-conserving real transitions.

Within the present single-mode model, the retained mode $g$
represents the common quantized gravitational degree of freedom
coupled to the two transverse oscillator modes of the detector.
This effective description allows us to investigate the
detector--field dynamics without introducing the full
gravitational spectral density. In particular, the leading
transition amplitudes involve the two pathways
\begin{equation}
|00;0_g\rangle
\longrightarrow
|02;1_g\rangle,
\qquad
|00;0_g\rangle
\longrightarrow
|20;1_g\rangle,
\end{equation}
which have equal magnitude and opposite sign in the present
model. Their relative phase is relevant to the coherence
structure of the reduced detector state obtained after tracing
over the gravitational degree of freedom. The resulting
quantum correlations and their entanglement properties,
however, must be assessed from the reduced density matrix
rather than inferred solely from the transition amplitudes.

A complete multimode treatment would require retaining the
full gravitational spectral density and summing or integrating
over all field modes. Such a treatment would determine explicitly
the frequency dependence of the coupling, the cumulative
dispersive contributions, and possible continuum-induced
modifications of the detector dynamics. Analogous issues arise
in the treatment of quantum systems coupled to radiation
continua, where resonant transitions and principal-value
contributions are separated within the Weisskopf--Wigner
framework \cite{WeisskopfWigner1930,PhysRevA.82.023818}.
The present single-mode calculation should therefore be
understood as an effective description of the selected
gravitational interaction rather than as a complete multimode
calculation.

Finally, it is important to emphasize that the single-mode
approximation and the rotating-wave approximation (RWA) are
logically independent. The single-mode approximation restricts
the gravitational field to one selected effective mode, whereas
the RWA neglects counter-rotating terms in the interaction
Hamiltonian. In the present analysis, the counter-rotating
creation terms are retained because they generate the
vacuum-induced transition amplitudes relevant to the
detector--detector coherence. The single-mode approximation
instead reflects the effective restriction of the gravitational
field to the selected mode, with the quantitative influence of
the omitted modes left for a future multimode investigation.

\subsection{Interaction-picture evolution and Magnus expansion}

Having established the effective single-mode description and clarified
the status of the omitted gravitational modes, we now investigate the
quantum dynamics of the detector--graviton system. We assume that the
two detectors are initially prepared in their ground states and that
the retained gravitational mode is initially in the vacuum,
\begin{equation}
 |\Psi(0)\rangle
 =
 (|0\rangle_1\otimes|0\rangle_2)_D
 \otimes|0_g\rangle
 \equiv
 |00;0_g\rangle_{t=0}.
 \label{xc}
\end{equation}

The interaction-picture evolution operator is formally written as
\begin{equation}
 \hat U_{\rm int}^{\hat\gamma}(t,0)
 =
 \hat T
 \exp
 \left[
 -
 \frac{i}{\hbar}
 \int_0^t
 dt'\,
 \hat H_{\rm int}^{I}(t')
 \right],
\label{Uint}
\end{equation}
where $\hat T$ denotes the time-ordering operator.

Since the detector--graviton interaction is weak within the
linearized gravitational-wave approximation, we retain terms up to
second order in the interaction Hamiltonian. A systematic perturbative
treatment that preserves the unitarity of the evolution operator at
each order is provided by the Magnus expansion
\cite{Blanes:2008xlr}.

Retaining contributions up to second order, the interaction-picture
evolution operator is
\begin{eqnarray}
 \hat U_{\rm int}^{\hat\gamma}(t,0)
 &\simeq&
 \exp
 \Bigg[
 -\frac{i}{\hbar}
 \int_0^t
 dt_1\,
 \hat H_{\rm int}^{I}(t_1)
 \nonumber\\
&&
-\frac{1}{2\hbar^2}
 \int_0^t
 dt_1
 \int_0^{t_1}
 dt_2
 \left[
 \hat H_{\rm int}^{I}(t_1),
 \hat H_{\rm int}^{I}(t_2)
 \right]
 \Bigg]
 \nonumber\\
&=&
1
-
\frac{i}{\hbar}
\int_0^t
dt_1\,
\hat H_{\rm int}^{I}(t_1)
\nonumber\\
&&
-
\frac{1}{2\hbar^2}
\int_0^t
dt_1
\int_0^{t_1}
dt_2
\left[
\hat H_{\rm int}^{I}(t_1),
\hat H_{\rm int}^{I}(t_2)
\right]
\nonumber\\
&&
-
\frac{1}{2\hbar^2}
\left(
\int_0^t
dt_1\,
\hat H_{\rm int}^{I}(t_1)
\right)^2
+
\mathcal O(C_\gamma^3).
\label{teo}
\end{eqnarray}

The second-order contribution is governed by the commutator of
interaction Hamiltonians evaluated at different times. Using the
bosonic commutation relations together with the free evolution of the
detector and gravitational operators, this commutator can be
evaluated analytically. It generates the leading quantum corrections
associated with the common gravitational degree of freedom. The
explicit calculation follows the procedure presented in the Appendix
of Ref.~\cite{Nandi:2022sjy}.

For the initial state above, the perturbative evolution can be written
as

\begin{eqnarray}
&&
|\Psi (t)\rangle_{f}
=
(\hat C_{00}^{(0)}
+
\hat C_{00}^{(2)})
|00;0\rangle
+
\hat C_{20}^{(1)}
|20;0\rangle
+
\hat C_{02}^{(1)}
|02;0\rangle
\nonumber\\
&&
+
\hat C_{40}^{(2)}
|40;0\rangle
+
\hat C_{04}^{(2)}
|04;0\rangle
+
\hat C_{22}^{(2)}
|22;0\rangle,
\label{finalstt}
\end{eqnarray}

where $\hat{C}_{00}^{(0)} = 1$ and
\begin{eqnarray}
\hat{C}_{00}^{(2)} &=& -\frac{4iC^2}{\hbar^2}{\int_{0}^{t}}dt_1{\int_{0}^{t_1}}dt_2\hat{\gamma}^I(t_1)\hat{\gamma}^I(t_2)\sin T
\nonumber
\\
&&+\frac{2C^2}{\hbar^2}{\int_{0}^{t}}dt_1{\int_{0}^{t_1}}dt_2[\hat{\gamma}^I(t_1),\hat{\gamma}^I(t_2)]{e^{T_-}}
\nonumber
\\
&&+\frac{2C^2}{\hbar^2}{\int_{0}^{t}}dt_2{\int_{0}^{t}}dt_1\hat{\gamma}^I(t_2)\hat{\gamma}^I(t_1){e^{T_-}}~;
\nonumber
\\
\hat{C}_{02}^{(1)}&=&\frac{\sqrt{2}iC}{\hbar}{\int_{0}^{t}}dt_1\hat{\gamma}^I(t_1){e^{2i\Omega_0 t_1}}~;
\nonumber
\\
\hat{C}_{20}^{(1)}&=&-\frac{\sqrt{2}iC}{\hbar}{\int_{0}^{t}}dt_1\hat{\gamma}^I(t_1){e^{2i\Omega_0 t_1}}~; 
\nonumber
\\
\hat{C}_{04}^{(2)} &=&-\frac{\sqrt{6}C^2}{\hbar^2}{\int_{0}^{t}}dt_1{\int_{0}^{t_1}}dt_2[\hat{\gamma}^I(t_1),\hat{\gamma}^I(t_2)]{e^{T_+}}
\nonumber
\\
&&-\frac{\sqrt{6}C^2}{\hbar^2}{\int_{0}^{t}}dt_2{\int_{0}^{t}}dt_1\hat{\gamma}^I(t_2)\hat{\gamma}^I(t_1){e^{T_+}}~;
\nonumber
\\
\hat{C}_{40}^{(2)}&=&-\frac{\sqrt{6}C^2}{\hbar^2}{\int_{0}^{t}}dt_1{\int_{0}^{t_1}}dt_2[\hat{\gamma}^I(t_1),\hat{\gamma}^I(t_2)]{e^{T_+}}
\nonumber
\\
&&-\frac{\sqrt{6}C^2}{\hbar^2}{\int_{0}^{t}}dt_2{\int_{0}^{t}}dt_1\hat{\gamma}^I(t_2)\hat{\gamma}^I(t_1){e^{T_+}}~; 
\nonumber
\\
{\hat{C}}_{22}^{(2)} &=& \frac{2C^2}{\hbar^2}{\int_{0}^{t}}dt_1{\int_{0}^{t_1}}dt_2[\hat{\gamma}^I(t_1),\hat{\gamma}^I(t_2)]{e^{T_+}}
\nonumber
\\
&+&\frac{2C^2}{\hbar^2}{\int_{0}^{t}}dt_2{\int_{0}^{t}}dt_1\hat{\gamma}^I(t_2)\hat{\gamma}^I(t_1){e^{T_+}}~.
\end{eqnarray}
Here we denote $C=i\hbar$, $T=2\Omega_0(t_1-t_2)$, $T_+=2i\Omega_0(t_1+t_2)$ and $T_-=2i\Omega_0(t_1-t_2)=iT$.

The coefficients
$\hat C_{20}^{(1)}$ and $\hat C_{02}^{(1)}$
describe first-order excitation of the individual detector modes,
whereas
$\hat C_{40}^{(2)}$,
$\hat C_{04}^{(2)}$,
and
$\hat C_{22}^{(2)}$
describe second-order processes generated by repeated interaction
with the quantized gravitational field. In particular,
$\hat C_{22}^{(2)}$ describes the simultaneous excitation of both
detector modes and provides the leading contribution associated with
correlations between the two detector subsystems.

It is important to distinguish the perturbative expansion of the
microscopic detector--gravity interaction from the perturbative
expansion of the subsequent quantum dynamics. Throughout this work,
the detector Hamiltonian is derived within leading-order linearized
gravity, where only terms linear in the metric perturbation are
retained. Consequently, the interaction Hamiltonian itself is
accurate only to first order in the gravitational-wave amplitude.
The higher-order coefficients appearing above do not arise from
introducing additional $O(h^2)$ interaction terms into the
Hamiltonian. Instead, they are generated by successive applications
of the same leading-order interaction Hamiltonian through the quantum
time-evolution operator.

The resulting global detector--graviton state evolves unitarily.
However, because both detector modes interact with the same
gravitational degree of freedom, the gravitational field can become
correlated with the detectors. Tracing out the gravitational mode can
therefore leave the two-detector subsystem in a nontrivial reduced
state. The existence of detector--graviton correlations does not by
itself establish detector--detector entanglement; genuine
detector--detector entanglement must be established by an independent
separability criterion applied to the reduced two-detector density
matrix.

Thus, the quantized gravitational field provides a common quantum
mediator between the two otherwise locally interacting detector
subsystems. Within the effective single-mode, narrow-band regime,
the selected gravitational mode captures the dominant dynamics
considered here, while the effects of omitted off-resonant modes are
understood as perturbative multimode corrections, including possible
dispersive shifts.

%%%%%%%%%%%%%%%%%%%%%%%%%%%%%%%%%%%%%%%%%%%%%%%%%%%%%%%%%%%%%%%
\section{Gravity-Induced Entanglement}
\label{sec:entanglement}
%%%%%%%%%%%%%%%%%%%%%%%%%%%%%%%%%%%%%%%%%%%%%%%%%%%%%%%%%%%%%%%

We now investigate whether the quantized gravitational field can generate
genuine quantum correlations between the two detector subsystems. The
analysis proceeds in two complementary stages. We first identify the
leading mechanism responsible for detector--detector entanglement by
restricting the reduced detector state to the minimal excitation sector.
We then return to the complete detector subspace populated by the
second-order perturbative dynamics in order to characterize the full
reduced detector state and its mixedness.

%%%%%%%%%%%%%%%%%%%%%%%%%%%%%%%%%%%%%%%%%%%%%%%%%%%%%%%%%%%%%%%
\subsection{Classical Stochastic Gravitational Waves}
%%%%%%%%%%%%%%%%%%%%%%%%%%%%%%%%%%%%%%%%%%%%%%%%%%%%%%%%%%%%%%%

Before considering the quantized gravitational field, it is useful to
establish the corresponding classical reference case \cite{PhysRevD.101.125018}. Suppose that the
gravitational-wave perturbation $h_{ij}^{\rm TT}(t)$ is treated as a
prescribed classical field. For a given realization of the classical
waveform, the detector interaction Hamiltonian can be written as
\begin{equation}
\hat H_{\rm int}^{\rm cl}(t)
=
\hat H_{{\rm int},1}^{\rm cl}(t)
+
\hat H_{{\rm int},2}^{\rm cl}(t),
\end{equation}
where each term acts only on the corresponding detector Hilbert space.
Since the two detector operators act on different subsystems,
\begin{equation}
\left[
\hat H_{{\rm int},1}^{\rm cl}(t),
\hat H_{{\rm int},2}^{\rm cl}(t')
\right]
=
0,
\end{equation}
and the corresponding interaction-picture evolution factorizes,
\begin{equation}
\hat U_{\rm cl}(t)
=
\hat U_1^{\rm cl}(t)
\otimes
\hat U_2^{\rm cl}(t).
\label{Uclassical}
\end{equation}

Consequently, an initially separable detector state,
\begin{equation}
\rho_{12}(0)
=
\rho_1(0)\otimes\rho_2(0),
\end{equation}
remains separable under the evolution generated by any fixed classical
gravitational-wave realization,
\begin{equation}
\rho_{12}^{\rm cl}(t)
=
\rho_1^{\rm cl}(t)
\otimes
\rho_2^{\rm cl}(t).
\end{equation}
If the gravitational-wave signal is stochastic, the experimentally
relevant state is obtained by averaging over the classical ensemble,
\begin{equation}
\rho_{12}^{\rm cl}
=
\int d\lambda\,
p(\lambda)\,
\rho_1^{(\lambda)}
\otimes
\rho_2^{(\lambda)},
\label{classicalensemble}
\end{equation}
where $\lambda$ labels the classical field realization. Equation
(\ref{classicalensemble}) is a convex mixture of product states and is
therefore separable \cite{PhysRevD.101.125018}.

Thus, a classical stochastic gravitational wave can generate local
excitation and mixedness after ensemble averaging, but it does not
generate genuine detector--detector entanglement through the local
interaction considered here. This provides the classical reference
against which the quantized gravitational-field description can be
distinguished.

%%%%%%%%%%%%%%%%%%%%%%%%%%%%%%%%%%%%%%%%%%%%%%%%%%%%%%%%%%%%%%%
\subsection{Quantized Gravitational Field and Reduced Detector State}
%%%%%%%%%%%%%%%%%%%%%%%%%%%%%%%%%%%%%%%%%%%%%%%%%%%%%%%%%%%%%%%

We now treat the gravitational field as a quantum degree of freedom.
The two detector subsystems interact locally with the same quantized
gravitational mode. Starting from an initially separable state of the
detectors and the gravitational field, the composite detector--graviton
system evolves unitarily according to the interaction-picture evolution
derived in the preceding section.

For the initial state
\begin{equation}
|\Psi(0)\rangle
=
|00\rangle_D\otimes|0_G\rangle,
\end{equation}
the perturbative evolution generates detector excitations correlated
with the gravitational field. At the perturbative order considered
here, the detector sector is populated by
\begin{equation}
|00\rangle,\qquad
|02\rangle,\qquad
|20\rangle,\qquad
|40\rangle,\qquad
|04\rangle,\qquad
|22\rangle.
\end{equation}
We therefore define the complete detector subspace populated by the
second-order dynamics as
\begin{equation}
\mathcal H_{\rm full}
=
{\rm span}
\left\{
|00\rangle,
|02\rangle,
|20\rangle,
|40\rangle,
|04\rangle,
|22\rangle
\right\}.
\label{Hfull}
\end{equation}
Here $\mathcal H_{\rm full}$ denotes the detector subspace populated
at the perturbative order considered, rather than the full
infinite-dimensional Hilbert space of the two oscillators.

The density operator of the combined detector--graviton system is
\begin{equation}
\hat\rho_f
=
|\Psi(t)\rangle_f\,{}_f\langle\Psi(t)|.
\label{FullDensity}
\end{equation}
Because the transition amplitudes appearing in Eq.~(\ref{finalstt})
are operators acting on the graviton Hilbert space, $\hat\rho_f$
contains products of operator-valued coefficients. The state relevant
to measurements performed only on the detector subsystem is therefore
obtained by tracing over the gravitational degrees of freedom,
\begin{equation}
\hat\rho_{12}
=
{\rm Tr}_G
\left(
\hat\rho_f
\right).
\label{ReducedDensity}
\end{equation}

To make the role of this partial trace explicit, consider the
interference contribution between the two first-order detector
excitation pathways,
\begin{equation}
\hat C_{02}^{(1)}
|02;0_G\rangle
\langle20;0_G|
\left(
\hat C_{20}^{(1)}
\right)^\dagger .
\end{equation}
Using
\begin{equation}
|02;0_G\rangle
=
|02\rangle\otimes|0_G\rangle,
\end{equation}
this contribution becomes
\begin{equation}
|02\rangle
\langle20|
\otimes
\hat C_{02}^{(1)}
|0_G\rangle
\langle0_G|
\left(
\hat C_{20}^{(1)}
\right)^\dagger .
\end{equation}
Taking the trace over the gravitational Hilbert space gives
\begin{equation}
{\rm Tr}_G
\left[
\hat C_{02}^{(1)}
|0_G\rangle
\langle0_G|
\left(
\hat C_{20}^{(1)}
\right)^\dagger
\right]
=
\langle0_G|
\left(
\hat C_{20}^{(1)}
\right)^\dagger
\hat C_{02}^{(1)}
|0_G\rangle.
\end{equation}
Consequently, the corresponding contribution to the reduced detector
density operator is
\begin{equation}
\langle0_G|
\left(
\hat C_{20}^{(1)}
\right)^\dagger
\hat C_{02}^{(1)}
|0_G\rangle
|02\rangle
\langle20|.
\label{InterferenceReduced}
\end{equation}

This illustrates how the operator-valued gravitational transition
amplitudes enter the detector density matrix. Before the partial trace,
the transition amplitudes act on the gravitational Hilbert space.
After tracing over the gravitational degrees of freedom, their
products become expectation values in the gravitational state and
therefore appear as ordinary coefficients multiplying detector
operators. In particular, the off-diagonal term in
Eq.~(\ref{InterferenceReduced}) retains the coherence between the two
distinct detector excitation pathways.

Proceeding in the same manner for all contributions to
$\hat\rho_f$, the reduced detector density operator can be expressed
generically as
\begin{equation}
\hat\rho_{12}
=
\sum_{ab,cd}
M_{ab,cd}
|ab\rangle\langle cd|,
\label{GeneralDensity}
\end{equation}
where the matrix elements are determined by gravitational expectation
values of products of the corresponding transition amplitudes,
\begin{equation}
M_{ab,cd}
=
\langle0_G|
\left(
\hat C_{cd}
\right)^\dagger
\hat C_{ab}
|0_G\rangle,
\label{MatrixElements}
\end{equation}
with the identity and second-order vacuum contributions understood to
be included in the appropriate perturbative coefficients.

Equation~(\ref{GeneralDensity}) represents the complete reduced
detector state within $\mathcal H_{\rm full}$. Its diagonal elements
describe detector excitation probabilities, whereas its off-diagonal
elements describe coherences between different detector excitation
pathways. These coherences are particularly important for determining
whether the two detector subsystems can become genuinely entangled.

The central question is therefore whether $\hat\rho_{12}$ is separable
with respect to the detector bipartition
\begin{equation}
D_1|D_2.
\end{equation}
Although the detectors have no direct interaction, each detector
couples locally to the same quantized gravitational degree of freedom.
The gravitational field can consequently mediate correlations between
the two detector subsystems.

%%%%%%%%%%%%%%%%%%%%%%%%%%%%%%%%%%%%%%%%%%%%%%%%%%%%%%%%%%%%%%%

%%%%%%%%%%%%%%%%%%%%%%%%%%%%%%%%%%%%%%%%%%%%%%%%%%%%%%%%%%%%%%%
\subsection{Leading Mechanism for Detector--Detector Entanglement}
%%%%%%%%%%%%%%%%%%%%%%%%%%%%%%%%%%%%%%%%%%%%%%%%%%%%%%%%%%%%%%%

To gain physical insight into the mechanism responsible for the
emergence of detector entanglement, we first examine the leading
excitation pathways generated by the detector--graviton interaction.
Retaining the first-order transition amplitudes, the
detector--graviton state takes the form
\begin{equation}
\begin{aligned}
|\Psi(t)\rangle_f
={}&
|00\rangle\otimes|0_G\rangle
\\
&+
|02\rangle\otimes
\hat C_{02}^{(1)}|0_G\rangle
+
|20\rangle\otimes
\hat C_{20}^{(1)}|0_G\rangle
+
\mathcal O(C_\gamma^2),
\end{aligned}
\label{LeadingState}
\end{equation}
where the transition amplitudes are operators acting on the
gravitational-field Hilbert space. In particular, the interaction
Hamiltonian in Eq.~(\ref{eq:HI_final}) gives
\begin{equation}
\hat C_{20}^{(1)}
=
-\hat C_{02}^{(1)}.
\end{equation}
Defining
\begin{equation}
\hat C
\equiv
\hat C_{02}^{(1)}
=
-\hat C_{20}^{(1)},
\end{equation}
we can write Eq.~(\ref{LeadingState}) as
\begin{equation}
\begin{aligned}
|\Psi(t)\rangle_f
={}&
|00\rangle\otimes|0_G\rangle
\\
&+
\left(|02\rangle-|20\rangle\right)
\otimes\hat C|0_G\rangle
+
\mathcal O(C_\gamma^2).
\end{aligned}
\label{LeadingStateCompact}
\end{equation}

The two first-order excitation pathways correspond to excitation of
either detector subsystem. Owing to the tensorial structure of the
plus-polarized gravitational-wave interaction, their transition
amplitudes have equal magnitude and opposite sign. Thus, the
detector excitation sector is associated with the antisymmetric
combination
\begin{equation}
|\Psi^-_{12}\rangle
=
\frac{1}{\sqrt{2}}
\left(|02\rangle-|20\rangle\right),
\label{BellLike}
\end{equation}
which has the form of a Bell-like state of the two detector
subsystems. Equation~(\ref{LeadingStateCompact}) therefore provides
a useful amplitude-level description of the correlations generated
by the common gravitational degree of freedom. Equivalently, it can
be expressed as
\begin{equation}
\begin{aligned}
|\Psi(t)\rangle_f
={}&
|00\rangle\otimes|0_G\rangle
\\
&+
\sqrt{2}\,
|\Psi^-_{12}\rangle\otimes\hat C|0_G\rangle
+
\mathcal O(C_\gamma^2).
\end{aligned}
\label{LeadingStateBell}
\end{equation}

It is important, however, to distinguish this amplitude-level
structure from a demonstration of detector--detector entanglement.
The appearance of an antisymmetric excitation amplitude does not,
by itself, establish that the reduced detector state is
nonseparable. This must be determined from the reduced density
operator obtained after tracing out the gravitational degrees of
freedom. Furthermore, as will be shown below, the second-order
transition amplitude to the doubly excited state $|22\rangle$
generates an additional coherence with the detector vacuum. This
contribution must be retained when evaluating the partial
transpose.

%%%%%%%%%%%%%%%%%%%%%%%%%%%%%%%%%%%%%%%%%%%%%%%%%%%%%%%%%%%%%%%
\subsection{Leading Entanglement Sector}
%%%%%%%%%%%%%%%%%%%%%%%%%%%%%%%%%%%%%%%%%%%%%%%%%%%%%%%%%%%%%%%

We now project the complete reduced detector state onto the
four-dimensional sector
\begin{equation}
\mathcal H_{\rm proj}
=
{\rm span}
\left\{
|00\rangle,\,
|02\rangle,\,
|20\rangle,\,
|22\rangle
\right\}.
\label{Hproj}
\end{equation}
This sector contains both the first-order single-subsystem
excitation pathways and the second-order doubly excited state.
The latter is essential because its transition amplitude can
interfere with the zeroth-order vacuum amplitude after tracing
over the gravitational field.

Introducing the projector
\begin{equation}
P
=
|00\rangle\langle00|
+
|02\rangle\langle02|
+
|20\rangle\langle20|
+
|22\rangle\langle22|,
\end{equation}
the projected reduced density operator is
\begin{equation}
\hat\rho_{\rm proj}
=
P\hat\rho_{12}P.
\label{ProjectedDensity}
\end{equation}
As in the preceding analysis, this projection is used as a
diagnostic of entanglement in the lowest relevant excitation
sector. Contributions involving $|40\rangle$ and $|04\rangle$
are not included in this projected calculation; they remain
part of the complete detector dynamics.

Using Eq.~(\ref{GeneralDensity}), we write
\begin{equation}
\hat\rho_{\rm proj}
=
\sum_{ab,cd\in\mathcal H_{\rm proj}}
M_{ab,cd}|ab\rangle\langle cd|.
\label{ProjectedGeneral}
\end{equation}

We first consider the coherence between the two single-subsystem
excitation pathways. Defining
\begin{equation}
\Gamma
=
\langle0_G|
\hat C^\dagger\hat C
|0_G\rangle,
\label{GammaDef}
\end{equation}
and using
$\hat C_{20}^{(1)}=-\hat C_{02}^{(1)}$, we obtain
\begin{equation}
M_{02,20}
=
-\Gamma,
\qquad
M_{20,02}
=
-\Gamma.
\label{M0220}
\end{equation}
For the graviton vacuum, the explicit first-order transition
amplitudes give
\begin{equation}
\Gamma
=
2C_\gamma^2
\int_0^t dt_1
\int_0^t dt_2\,
e^{i(\omega_0+\omega_g)(t_1-t_2)},
\label{GammaIntegral}
\end{equation}
where $\omega_0=2\Omega_0$. Equivalently,
\begin{equation}
\Gamma
=
2C_\gamma^2
\left|
\int_0^t dt'\,
e^{i(\omega_0+\omega_g)t'}
\right|^2
\geq 0.
\label{GammaPositive}
\end{equation}
In particular, $\Gamma$ is of order $C_\gamma^2$.

We next retain the second-order amplitude
$\hat C_{22}^{(2)}$ appearing in Eq.~(\ref{finalstt}). Its
contribution to the reduced detector state gives rise to the
vacuum--doubly-excited coherence
\begin{equation}
z(t)
\equiv
\langle0_G|
\hat C_{22}^{(2)}(t)
|0_G\rangle.
\label{zDefinition}
\end{equation}
To evaluate this quantity, we use the second-order expression
for $\hat C_{22}^{(2)}$ in Eq.~(\ref{C22}) and the vacuum
correlation function
\begin{equation}
\langle0_G|
\hat\gamma_I(t_1)\hat\gamma_I(t_2)
|0_G\rangle
=
C_\gamma^2e^{-i\omega_g(t_1-t_2)}.
\label{VacuumCorrelation}
\end{equation}
Combining the commutator and full-square contributions in
Eq.~(\ref{C22}), the second-order amplitude can be expressed as
\begin{equation}
\hat C_{22}^{(2)}(t)
=
-4\int_0^t dt_1
\int_0^{t_1}dt_2\,
e^{i\omega_0(t_1+t_2)}
\hat\gamma_I(t_1)\hat\gamma_I(t_2).
\label{C22Dyson}
\end{equation}
Consequently,
\begin{equation}
z(t)
=
-4C_\gamma^2
\int_0^t dt_1
\int_0^{t_1}dt_2\,
e^{i(\omega_0-\omega_g)t_1}
e^{i(\omega_0+\omega_g)t_2}.
\label{zIntegral}
\end{equation}

Introducing
\begin{equation}
I(q,t)
\equiv
\int_0^t du\,e^{iqu}
=
\frac{e^{iqt}-1}{iq},
\qquad
I(0,t)=t,
\label{Idef}
\end{equation}
the ordered double integral can be evaluated to give
\begin{equation}
z(t)
=
-\frac{4C_\gamma^2}{i(\omega_0+\omega_g)}
\left[
I(2\omega_0,t)
-
I(\omega_0-\omega_g,t)
\right].
\label{zExplicit}
\end{equation}
Thus, the coherence $z(t)$ is also of order $C_\gamma^2$.

The projected density matrix must therefore retain both the
single-excitation coherence and the vacuum--doubly-excited
coherence. In the ordered basis
$\{|00\rangle,|02\rangle,|20\rangle,|22\rangle\}$, it takes
the form
\begin{equation}
\rho_{\rm proj}
=
\begin{pmatrix}
1-2\Gamma & 0 & 0 & z^*\\
0 & \Gamma & -\Gamma & 0\\
0 & -\Gamma & \Gamma & 0\\
z & 0 & 0 & p_{22}
\end{pmatrix}
+\text{higher-order contributions},
\label{rhoproj}
\end{equation}
where
\begin{equation}
p_{22}
=
\langle0_G|
\hat C_{22}^{(2)\dagger}
\hat C_{22}^{(2)}
|0_G\rangle
=
\mathcal O(C_\gamma^4).
\label{p22}
\end{equation}
Here the displayed diagonal entries and coherences are retained
at their respective leading nonvanishing orders. In particular,
the vacuum--doubly-excited coherence is of order $C_\gamma^2$,
whereas the population $p_{22}$ starts at order $C_\gamma^4$.
The projected matrix is normalized through the displayed
second order; a fully normalized expression at fourth order
would require the corresponding higher-order corrections.

We now apply the Peres--Horodecki positive-partial-transpose
(PPT) criterion to examine the entanglement between the
two detector subsystems
\cite{Horodecki:1996nc, Peres:1996dw}.
A bipartite state is said to be PPT if its partially
transposed density matrix is positive semidefinite,
whereas it is said to be NPT (negative partial transpose)
if the partial transpose possesses at least one negative
eigenvalue. An NPT state is necessarily entangled.

Taking the partial transpose with respect to subsystem $2$,
we obtain
\begin{equation}
\rho_{\rm proj}^{T_2}
=
\begin{pmatrix}
1-2\Gamma & 0 & 0 & -\Gamma\\
0 & \Gamma & z^* & 0\\
0 & z & \Gamma & 0\\
-\Gamma & 0 & 0 & p_{22}
\end{pmatrix}
+\text{higher-order contributions}.
\label{PTT}
\end{equation}
This matrix separates into two independent $2\times2$
blocks. In particular, the block acting on the subspace
spanned by $\{|02\rangle,|20\rangle\}$ is
\begin{equation}
M_{\rm PT}
=
\begin{pmatrix}
\Gamma & z^*\\
z & \Gamma
\end{pmatrix}.
\label{MPT}
\end{equation}
Its eigenvalues are
\begin{equation}
\lambda_\pm^{\rm PT}
=
\Gamma\pm |z|.
\label{PTeigenvalues}
\end{equation}
The smaller eigenvalue is therefore
\begin{equation}
\lambda_-^{\rm PT}
=
\Gamma-|z|.
\label{PTeigenvalueMinus}
\end{equation}
Consequently, the projected detector state is NPT at
leading order whenever
\begin{equation}
|z(t)|>\Gamma(t).
\label{NPTcondition}
\end{equation}
Since the projected state is defined on a
$2\times2$ detector Hilbert space, the PPT criterion is
also necessary and sufficient for its separability
\cite{Peres:1996dw}. Thus, within the projected
description, the condition in Eq.~(\ref{NPTcondition})
establishes entanglement between the two detector
subsystems.

Importantly, both $\Gamma$ and $z$ are of order
$C_\gamma^2$, so that the negative eigenvalue
$\lambda_-^{\rm PT}$ appears already at this order.
Its sign is therefore determined by the leading
perturbative contribution whenever the inequality in
Eq.~(\ref{NPTcondition}) is satisfied with a strict
margin. This avoids relying on a negative eigenvalue
of order $C_\gamma^4$ extracted from a density matrix
truncated at second order.

For completeness, the other partial-transpose block is
\begin{equation}
M_0
=
\begin{pmatrix}
1-2\Gamma & -\Gamma\\
-\Gamma & p_{22}
\end{pmatrix}.
\label{M0}
\end{equation}
Its determinant depends on the fourth-order population
$p_{22}$:
\begin{equation}
\det M_0
=
(1-2\Gamma)p_{22}-\Gamma^2.
\label{M0det}
\end{equation}
A conclusion based on the sign of this determinant requires
a consistent calculation through fourth order. We therefore
do not use this block to establish the leading-order
entanglement result.

The leading negativity associated with the NPT block is
\begin{equation}
\mathcal N_{\rm leading}
=
\max\{0,\,|z|-\Gamma\},
\label{negativity}
\end{equation}
up to higher-order corrections, whenever the leading-order
eigenvalue is separated from zero. At points where
$|z|=\Gamma$, or in parameter regions where the difference
is comparable to neglected corrections, higher-order terms
must be retained before drawing a conclusion.

The condition in Eq.~(\ref{NPTcondition}) can be expressed
in a dimensionless form. Defining
\begin{equation}
x=\omega_0t,
\qquad
r=\frac{\omega_g}{\omega_0},
\qquad
F(y)=\frac{e^{iy}-1}{iy},
\qquad F(0)=1,
\label{DimensionlessVariables}
\end{equation}
we find
\begin{equation}
\Gamma
=
2C_\gamma^2t^2
\left|F((1+r)x)\right|^2
\end{equation}
and
\begin{equation}
|z|
=
\frac{4C_\gamma^2t}{(1+r)\omega_0}
\left|F(2x)-F((1-r)x)\right|.
\end{equation}
Hence, the leading-order NPT condition becomes
\begin{equation}
\frac{|z|}{\Gamma}
=
\frac{
2\left|F(2x)-F((1-r)x)\right|
}{
(1+r)x\left|F((1+r)x)\right|^2
}
>1.
\label{DimensionlessNPT}
\end{equation}
This criterion makes explicit that the existence of the
leading-order NPT signature depends on the dimensionless
interaction time and the ratio of the gravitational-mode
frequency to the detector transition frequency. It is
therefore not a parameter-independent conclusion; the
inequality must be checked in the parameter regime under
consideration.

To examine the dependence of the NPT condition on both
the interaction time and the gravitational-mode frequency,
we introduce the bounded quantity
\begin{equation}
\mathcal{W}(x;r)
=
\frac{|z|-\Gamma}{|z|+\Gamma},
\label{BoundedNPTWitness}
\end{equation}
whenever $|z|+\Gamma\neq0$. By construction,
$-1\leq\mathcal{W}\leq1$, and the leading-order NPT
condition is equivalent to
\begin{equation}
\mathcal{W}(x;r)>0.
\end{equation}
Thus, positive values of $\mathcal{W}$ identify parameter
regions in which the considered partial-transpose block
has a negative eigenvalue at leading order, whereas
negative values indicate that this block does not
establish NPT. The quantity $\mathcal{W}$ is used here
only as a bounded diagnostic of the NPT condition and
not as an entanglement measure.

\begin{figure}[t]
    \centering
    \includegraphics[width=0.85\textwidth]
    {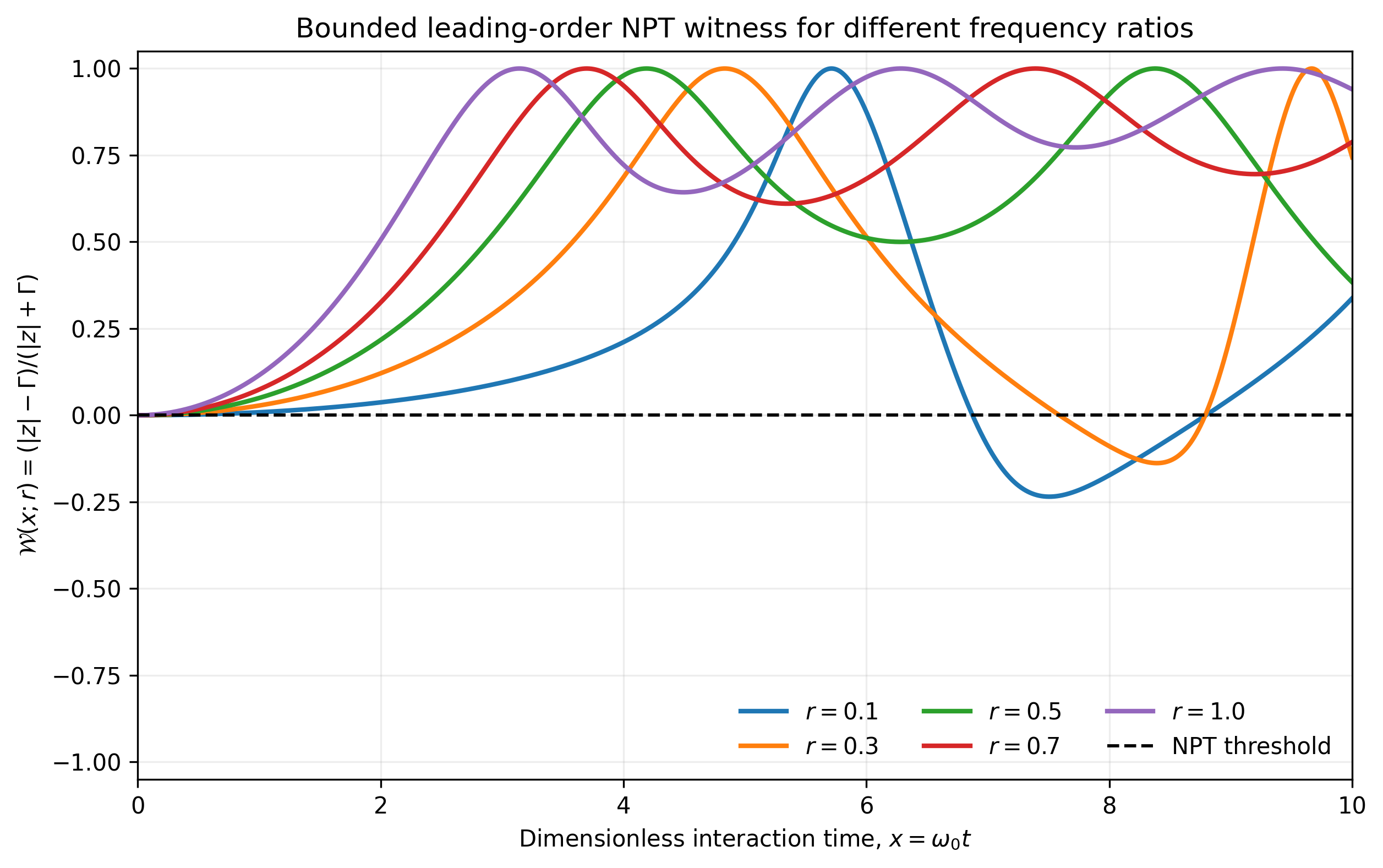}
    \caption{
    Bounded diagnostic $\mathcal{W}(x;r)
    =(|z|-\Gamma)/(|z|+\Gamma)$ as a function of the
    dimensionless interaction time $x=\omega_0t$ for
    different gravitational-mode frequency ratios
    $r=\omega_g/\omega_0=0.1,\,0.3,\,0.5,\,0.7,$ and $1.0$.
    The horizontal dashed line at $\mathcal{W}=0$
    marks the leading-order NPT threshold. Positive
    values correspond to a negative eigenvalue of the
    considered partial-transpose block, whereas negative
    values indicate that this block does not establish
    NPT. Unlike the ratio $|z|/\Gamma$, this bounded
    representation remains finite at zeros of $\Gamma$
    whenever $|z|\neq0$.
    }
    \label{fig:NPT_family}
\end{figure}

Figure~\ref{fig:NPT_family} displays $\mathcal{W}(x;r)$
as a function of the dimensionless interaction time for
several values of the frequency ratio,
$r=0.1,\,0.3,\,0.5,\,0.7,$ and $1.0$. The horizontal
dashed line at $\mathcal{W}=0$ separates the regions
satisfying and not establishing the leading-order NPT
condition. The family of curves illustrates how the
time intervals over which the nonlocal coherence
dominates over the local excitation contribution depend
on the frequency of the retained gravitational mode.
In particular, the comparison shows that the leading-order
NPT condition is sensitive to both the interaction time
and the gravitational-mode frequency.

As a representative point on the curve corresponding
to $r=0.3$ in Fig.~\ref{fig:NPT_family}, consider
\begin{equation}
r=0.3,
\qquad
x=3,
\end{equation}
corresponding to $\omega_g=0.3\omega_0$ and
$t=3/\omega_0$. Here, $r$ specifies the frequency of the
retained gravitational mode in the single-mode approximation,
while the initial gravitational state is the vacuum
$|0_g\rangle$, as specified in Eq.~(\ref{xc}).
Thus, the one-graviton component entering the first-order
transition amplitudes is generated dynamically by the
interaction, rather than being present in the initial state.
For these parameter values, Eqs.~(\ref{GammaPositive})
and~(\ref{zExplicit}) give
\begin{equation}
\Gamma
\simeq
4.085\,\frac{C_\gamma^2}{\omega_0^2},
\qquad
|z|
\simeq
7.797\,\frac{C_\gamma^2}{\omega_0^2}.
\end{equation}
Consequently, the smaller eigenvalue of the relevant
partial-transpose block is
\begin{equation}
\lambda_-^{\rm PT}
=
\Gamma-|z|
\simeq
-3.712\,\frac{C_\gamma^2}{\omega_0^2}<0.
\label{NumericalPTEigenvalue}
\end{equation}
The negative sign establishes that the leading-order
NPT condition is satisfied for this choice of parameters.
For the same parameter values, the bounded diagnostic
takes the value
\begin{equation}
\mathcal{W}(3;0.3)
=
\frac{7.797-4.085}{7.797+4.085}
\simeq 0.312.
\end{equation}
The perturbative interpretation, however, requires
$\Gamma\ll 1$ and $|z|\ll 1$, which in this example
amounts to
\begin{equation}
7.797\,\frac{C_\gamma^2}{\omega_0^2}\ll 1.
\end{equation}
This example illustrates a parameter regime in which
the leading-order entanglement witness is negative,
with the negative eigenvalue appearing at order
$C_\gamma^2$. It is a result within the vacuum-initiated,
single-mode approximation and should not be interpreted
as a parameter-independent conclusion, a demonstration
of experimental observability, or a result established
for the full multimode gravitational field.

The physical mechanism can therefore be summarized as
\begin{equation}
\text{detector}_1
\longleftrightarrow
\text{quantized gravitational field}
\longleftrightarrow
\text{detector}_2,
\end{equation}
with the entanglement witness arising from the combined
effect of the single-excitation coherence and the
vacuum--doubly-excited coherence generated by the common
gravitational degree of freedom.

%%%%%%%%%%%%%%%%%%%%%%%%%%%%%%%%%%%%%%%%%%%%%%%%%%%%%%%%%%%%%%%
\subsection{Return to the Complete Second-Order Detector State}
%%%%%%%%%%%%%%%%%%%%%%%%%%%%%%%%%%%%%%%%%%%%%%%%%%%%%%%%%%%%%%%

The four-dimensional calculation above serves a deliberately specific
purpose: it isolates the minimal sector containing the leading
$|02\rangle$--$|20\rangle$ coherence and makes the origin of the PPT
violation analytically transparent. It is not intended to replace the
complete second-order detector state.

Having established the existence of detector--detector entanglement in
this minimal sector, we therefore return to the complete six-dimensional
detector space,
\begin{equation}
\mathcal H_{\rm full}
=
{\rm span}
\left\{
|00\rangle,
|02\rangle,
|20\rangle,
|40\rangle,
|04\rangle,
|22\rangle
\right\}.
\end{equation}

This step is necessary because the second-order perturbative dynamics
populates the additional states $|40\rangle$, $|04\rangle$, and
$|22\rangle$, together with their associated populations and
coherences. These terms are not required to identify the leading
entanglement mechanism, but they must be retained when evaluating the
complete reduced-state dynamics at the same perturbative order.

Thus, the two calculations have distinct purposes. The projected
four-dimensional calculation identifies and establishes the leading
nonseparable detector correlation, while the complete
six-dimensional calculation retains all states populated at second
order and provides the appropriate reduced state for the subsequent
analysis of local entropy, purity, and the full detector dynamics.

%%%%%%%%%%%%%%%%%%%%%%%%%%%%%%%%%%%%%%%%%%%%%%%%%%%%%%%%%%%%%%%
\subsection{Entanglement Entropy and Purity}
%%%%%%%%%%%%%%%%%%%%%%%%%%%%%%%%%%%%%%%%%%%%%%%%%%%%%%%%%%%%%%%

Having established genuine detector--detector entanglement through the
PPT criterion, we next characterize the mixedness of an individual
detector using the complete second-order reduced state. This provides
information complementary to the negativity: whereas the negativity
tests whether the two detectors share genuine nonseparable
correlations, the entropy and purity quantify how strongly an
individual detector is correlated with degrees of freedom that have
been traced out.

The reduced density matrix of detector $1$ is obtained by tracing over
detector $2$,
\begin{equation}
\hat\rho_1(t)
=
{\rm Tr}_2
\left[
\hat\rho_{12}(t)
\right]
=
\sum_{n_2}
\langle n_2|
\hat\rho_{12}(t)
|n_2\rangle.
\label{rho1}
\end{equation}

Substituting the complete second-order reduced density matrix gives
\begin{eqnarray}
\hat\rho_1
&=&
K_{00}|0\rangle\langle0|
+
K_{22}|2\rangle\langle2|
\nonumber\\
&&+
K_{40}|4\rangle\langle0|
+
K_{40}^{*}|0\rangle\langle4|,
\label{redden}
\end{eqnarray}
which acts on the local detector subspace
\begin{equation}
\mathcal H_1
=
{\rm span}
\left\{
|0\rangle,
|2\rangle,
|4\rangle
\right\}.
\end{equation}

The coefficients are determined by the perturbative transition
amplitudes. In particular,
\begin{align}
K_{00}
&=
\Big\langle
1
+
\hat C_{00}^{(2)}
+
\hat C_{00}^{(2)\dagger}
+
\hat C_{02}^{(1)\dagger}
\hat C_{02}^{(1)}
\Big\rangle_G,
\\
K_{22}
&=
\Big\langle
\hat C_{20}^{(1)\dagger}
\hat C_{20}^{(1)}
\Big\rangle_G,
\\
K_{40}
&=
\Big\langle
\hat C_{40}^{(2)}
\Big\rangle_G.
\end{align}

To second order in the detector--graviton coupling, the relevant
eigenvalues of the reduced detector state are
\begin{equation}
K_{00}
=
1-K^{(2)},
\qquad
K_{22}
=
K^{(2)},
\label{Keigenvalues}
\end{equation}
where
\begin{equation}
K^{(2)}
=
2
\int_0^t
dt_1
\int_0^t
dt_2\,
e^{T_-}
\left\langle
\hat\gamma^I(t_2)
\hat\gamma^I(t_1)
\right\rangle_G .
\label{vac}
\end{equation}

For the vacuum graviton state and the interaction-picture mode
expansion derived above, this quantity becomes
\begin{equation}
K^{(2)}(t)
=
\frac{8C_\gamma^2}{\Omega^2}
\sin^2
\left(
\frac{\Omega t}{2}
\right),
\qquad
\Omega=2\Omega_0+\omega_g.
\label{B1}
\end{equation}
Thus,
\begin{equation}
K^{(2)}(t)\geq0.
\end{equation}

The von Neumann entropy of the reduced detector state is
\begin{equation}
S(t)
=
-
{\rm Tr}
\left[
\hat\rho_1(t)\ln\hat\rho_1(t)
\right].
\label{entropy}
\end{equation}
Using Eq.~(\ref{Keigenvalues}), this becomes
\begin{equation}
S(t)
=
-
K_{00}\ln K_{00}
-
K_{22}\ln K_{22},
\end{equation}
or
\begin{equation}
S(t)
=
-
\left[1-K^{(2)}(t)\right]
\ln\left[1-K^{(2)}(t)\right]
-
K^{(2)}(t)\ln K^{(2)}(t).
\label{entropyK}
\end{equation}

For $K^{(2)}\ll1$, the leading behavior is
\begin{equation}
S(t)
\simeq
K^{(2)}(t)
-
K^{(2)}(t)\ln K^{(2)}(t)
+
\mathcal O\!\left([K^{(2)}]^2\right).
\label{entropyApprox}
\end{equation}
The entropy therefore vanishes initially and becomes positive whenever
the detector develops correlations with degrees of freedom that are
not retained in its reduced state.

The oscillatory dependence
\begin{equation}
K^{(2)}(t)
\propto
\sin^2
\left(
\frac{\Omega t}{2}
\right)
\end{equation}
reflects the coherent time dependence of the detector--graviton
interaction. In particular, the resulting mixedness is not
irreversible decoherence: the complete detector--graviton evolution
remains unitary, and the apparent loss of local purity arises because
part of the global quantum state resides in degrees of freedom that
have been traced out.

An equivalent characterization is provided by the purity,
\begin{equation}
P(t)
=
{\rm Tr}
\left[
\hat\rho_1^2(t)
\right].
\label{purity}
\end{equation}
Using Eq.~(\ref{Keigenvalues}),
\begin{equation}
P(t)
=
K_{00}^2+K_{22}^2
=
\left[1-K^{(2)}(t)\right]^2
+
\left[K^{(2)}(t)\right]^2.
\end{equation}
Since $K^{(2)}=O(C_\gamma^2)$, the second term contributes only at
fourth order in the detector--graviton coupling. Therefore, to second
order,
\begin{equation}
P(t)
=
1
-
2K^{(2)}(t)
+
\mathcal O(C_\gamma^4).
\label{Purity}
\end{equation}
Initially,
\begin{equation}
P(0)=1,
\end{equation}
as expected for the initial pure state. Whenever
$K^{(2)}(t)>0$, the purity decreases below unity, indicating that
detector $1$ has become correlated with the degrees of freedom that
have been traced out.

The local entropy and purity therefore provide an information-theoretic
description of the redistribution of quantum correlations within the
combined detector--graviton system. A reduction in the purity of one
detector indicates that information initially localized in that
subsystem has become distributed among the remaining degrees of
freedom. In the present setup, both detectors interact with the same
gravitational mode, so this local mixedness reflects the development of
correlations within the combined detector--detector--graviton system.

It is important, however, to distinguish this local mixedness from
genuine detector--detector entanglement. A nonzero entropy or a reduced
purity does not by itself constitute a distinctive signature of a
quantized gravitational field, since a classical stochastic
gravitational-wave background can also produce mixed detector states
through ensemble averaging. The distinctive result of the present
analysis is instead the non-vanishing negativity of the two-detector
reduced state. The entropy and purity characterize the mixedness
associated with correlations with traced-out degrees of freedom,
whereas the negativity establishes genuine nonseparability between the
two detector subsystems.

Taken together, the two levels of analysis provide a complementary
description of the gravity-mediated quantum dynamics. The
four-dimensional projected calculation isolates the leading
$|02\rangle$--$|20\rangle$ coherence and demonstrates explicitly,
through the PPT criterion, that the detector subsystems become
entangled. Returning to the complete second-order detector subspace
then incorporates all perturbatively populated detector states and
allows the associated local mixedness to be quantified through the
von Neumann entropy and purity. The resulting picture is that the
quantized gravitational field acts as a common quantum mediator:
quantum information is distributed among the two detectors and the
gravitational degree of freedom, while a nonseparable component of
that information remains accessible in the two-detector subsystem.

\section{Discussion}
%%%%%%%%%%%%%%%%%%%%%%%%%%%%%%%%%%%%%%%%%%%%%%%%%%%%%%%%%%%%%%%

The results obtained in the preceding sections provide an operational
framework for investigating whether a propagating gravitational-wave (GW)
field can mediate quantum correlations between the two independent
transverse oscillator modes of a single localized detector system. Starting
from the geodesic-deviation equation in linearized General Relativity, we
constructed the detector--GW interaction and treated the gravitational-wave
field either as a classical stochastic field or as a quantized dynamical
degree of freedom. In the quantum description, the two oscillator modes,
$D_x$ and $D_y$, couple locally to the same gravitational mode, while no
direct coupling between the two modes is introduced. The resulting dynamics
therefore allows us to investigate whether the common propagating
gravitational field can generate quantum correlations across the
$D_x|D_y$ mode bipartition of the detector.

A central result of our analysis is that the quantized gravitational-wave
field can generate genuine entanglement across the two oscillator modes.
The leading excitation sector contains coherence between the two
indistinguishable excitation pathways, $|02\rangle$ and $|20\rangle$, whose
relative sign produces the characteristic antisymmetric combination of the
two transverse modes. After tracing over the gravitational degrees of
freedom, the resulting reduced detector state violates the
positive-partial-transpose criterion and possesses nonzero negativity.
Thus, although no direct coupling between $D_x$ and $D_y$ is introduced,
their local coupling to the same quantized gravitational field can establish
a nonseparable state across the $D_x|D_y$ mode bipartition.

The entropy and purity provide a complementary characterization of this
process. The complete detector--graviton state remains globally pure under
the unitary evolution, whereas tracing over the gravitational degrees of
freedom generally leaves the two-mode detector state in a mixed state. Such
mixedness should not, by itself, be interpreted as evidence for a quantum
gravitational mediator, since classical stochastic gravitational-wave
fluctuations can also reduce detector purity. The relevant distinction in
the present framework is therefore the entanglement across the $D_x|D_y$
mode bipartition. For a prescribed classical GW realization, the evolution
of the two oscillator modes remains factorized, and stochastic averaging
produces a convex mixture of product states. The resulting state may be
mixed but remains separable with respect to the $D_x|D_y$ bipartition. By
contrast, when the gravitational-wave field is quantized, it constitutes an
additional quantum subsystem that can retain coherence between different
interaction pathways and thereby mediate nonseparable correlations between
the two modes. Within the assumptions of the present independent-mediator
framework, the appearance of entanglement across the $D_x|D_y$ bipartition
therefore distinguishes the quantized gravitational-wave description from
the corresponding classical stochastic description.

It is useful to place the present results in the context of
recent analyses of quantum signatures in weakly coupled
wave systems. In particular, Bao et al.~\cite{Bao:2026prd}
have identified important limitations on the detection
of nonclassical properties of weakly coupled waves,
including gravitational radiation. Their analysis
emphasizes two distinct obstructions: the effective
coarse-graining of field modes in realistic detection
processes, which can wash out nonclassical features,
and the additional suppression of intrinsically quantum
observables by powers of the weak interaction strength.
Importantly, these suppressions are not generally
compensated by increasing the occupation number of the
incident field. This highlights the distinction between
occupation-enhanced classical responses and genuinely
quantum signatures of the field.

The interaction considered here, however, has a
different structure from the linear mode-mixing
interaction underlying the analysis of
Ref.~\cite{Bao:2026prd}. In our model, the gravitational
field couples to quadratic combinations of detector
creation and annihilation operators, thereby inducing
two-quantum detector transitions. Consequently, the
relevant transition amplitudes, resonance conditions,
and reduced-state correlations must be evaluated within
this specific interaction model. The suppression
arguments of Ref.~\cite{Bao:2026prd} therefore provide an
important general caution concerning experimental
observability, but their quantitative implications for
the present mechanism cannot be inferred without a
dedicated analysis.

In particular, the present calculation establishes a
perturbative mechanism through which a quantized
gravitational mode can generate entanglement between
the detector subsystems. It does not establish that
these correlations are readily observable under
realistic experimental conditions. Their robustness
and observability will depend on the effective
gravitational coupling, the initial quantum state of
the gravitational field, the detector coherence
properties, the mode structure of the gravitational
field, and additional environmental channels.
A quantitative assessment will therefore require
an analysis that incorporates realistic detector
couplings, mode coarse-graining, and competing
sources of decoherence.

The conceptual significance of this result is closely connected with the
long-standing question of whether a classical gravitational field can
consistently participate in coherent quantum dynamics. The
Eppley--Hannah thought experiment~\cite{Eppley:1977emg} highlighted this
issue by considering the interaction between a classical gravitational
field and a quantum system. Their argument has subsequently been debated
because of its idealized assumptions concerning measurement, backreaction,
and the formulation of a consistent classical--quantum interaction. The
present work does not attempt to resolve that debate directly. Instead, it
addresses the question operationally by considering an independently
propagating gravitational-wave field and comparing its classical stochastic
and quantum descriptions.

Our framework should also be distinguished from the gravity-mediated
entanglement proposals of Bose \emph{et al.} and Marletto and
Vedral~\cite{Bose,Marletto:2017kzi}. In those proposals, the quantum systems
participating in the experiment also act as gravitational sources, and their
mutual gravitational interaction is used to probe the quantum nature of the
mediator. In the present work, by contrast, the detector subsystems are
localized quantum probes of an independently propagating gravitational-wave
field. The gravitational-wave field is therefore separated conceptually
from the detector sources, allowing the role of propagating gravitational
radiation as a mediator to be studied directly.

More generally, the distinction between a physical mediator and an
effective interaction is important in interpreting gravity-mediated
entanglement. The observation of entanglement generated by an effective
gravitational interaction does not, by itself, establish the quantum nature
of the underlying mediator, since field degrees of freedom can be integrated
out to produce an effective nonlocal interaction. More recent analyses have
also shown why the identity of the mediator and the available communication
channels must be specified carefully. Aziz and Howl~\cite{AzizHowl}
demonstrated that entanglement can arise in a classical gravitational
background when quantum matter provides additional communication channels
through virtual propagators. Marletto \emph{et al.}~\cite{Marletto2025}
emphasized that such scenarios need not contradict information-theoretic
no-go arguments because the relevant quantum information can be carried by
additional quantum matter degrees of freedom rather than by the classical
gravitational field itself. Di Biagio~\cite{DiBiagio:2025twt} further
clarified that the classical-mediator no-go result depends on the
simultaneous assumptions that the mediator is an independent physical
subsystem, that communication between the quantum systems occurs
exclusively through local interactions with that mediator, and that the
mediator itself is classical. Boulle and Franzmann~\cite{tgjx-y7hl} have
additionally emphasized that subsystem decomposition, tensor-product
factorization, and gravitational dressing require particular care in gauge
theories.

These distinctions are important for interpreting the present result. Our
analysis is not intended as an exception to existing information-theoretic
no-go arguments. Rather, the model explicitly specifies the assumptions
under which the comparison is made. The propagating gravitational-wave field
is treated as an independent physical subsystem, the detector subsystems
communicate only through their local coupling to this common field, and the
classical model contains no additional quantum communication channel. Under
these assumptions, the classical stochastic description produces no
detector entanglement, whereas quantizing the same propagating
gravitational-wave degree of freedom permits the field to mediate quantum
correlations. The present calculation therefore provides a dynamical
realization of this information-theoretic distinction in the setting of
propagating gravitational radiation.

The present framework therefore occupies a distinct position within the
gravity-mediated-entanglement programme. The key feature is the separation
between the gravitational field and the quantum probes: the detectors do not
generate the gravitational field responsible for their correlations.
Instead, they locally probe a common propagating gravitational-wave mode.
This setup allows the more specific operational question to be addressed:
whether the quantum character of a propagating gravitational-wave field can
be revealed through the correlations that it establishes between otherwise
noninteracting quantum detector systems.

\begin{table*}[t]
\centering
\caption{Comparison between representative approaches to
gravity-mediated quantum correlations and the present framework.}
\label{tab:comparison}
\renewcommand{\arraystretch}{1.15}
\begin{tabular}{p{2.7cm} p{2.5cm} p{3.0cm} p{4.2cm}}
\hline
\textbf{Approach}
&
\textbf{Mediator}
&
\textbf{Role of quantum systems}
&
\textbf{Main distinction from the present work}
\\
\hline

Eppley--Hannah~\cite{Eppley:1977emg}
&
Classical gravitational field
&
Quantum matter interacting with gravity
&
Foundational question concerning the consistency of
classical gravity coupled to quantum matter
\\

Bose \emph{et al.} and
Marletto--Vedral~\cite{Bose,Marletto:2017kzi}
&
Gravitational interaction
&
Quantum systems also act as gravitational sources
&
Gravity-mediated entanglement is used as an operational
probe of the mediator's quantum nature
\\

Aziz--Howl~\cite{AzizHowl}
&
Classical gravitational background
&
Quantum matter fields
&
Additional quantum matter channels can transmit
correlations
\\

Di Biagio~\cite{DiBiagio:2025twt}
&
Classical mediator under specified assumptions
&
Quantum subsystems
&
Clarifies the assumptions required for the
information-theoretic no-go result
\\

Boulle--Franzmann~\cite{tgjx-y7hl}
&
Gravitational degrees of freedom
&
Quantum subsystems
&
Emphasizes the subtleties of subsystem decomposition,
factorization, and gravitational dressing
\\

Present work
&
Propagating GW field
&
Quantum detector probes
&
The detectors are not gravitational sources;
the propagating GW field is the common mediator and is
treated either classically or quantum mechanically
\\

\hline
\end{tabular}
\end{table*}

Our analysis is also complementary to recent approaches that investigate
quantum signatures of gravitational radiation through gravitational
fluctuations, graviton--matter energy exchange, or direct graviton detection.
Cho and Hu studied quantum fluctuations of the gravitational field and their
manifestation as stochastic forces and open-system effects on geodesic
separation~\cite{PhysRevD.105.086004}. Their analysis emphasizes the role of
gravitational fluctuations as an environment and the distinction between
quantum and stochastic descriptions at the level of noise and dissipation.
In the present work, by contrast, the gravitational degree of freedom is
retained explicitly as a dynamical quantum subsystem, and the observable of
interest is the nonseparable correlation generated between two localized
quantum detector systems. The two perspectives are therefore complementary:
one characterizes the influence of gravitational quantum fluctuations on a
quantum system, while the other examines correlations established by the
gravitational field between separate quantum systems.

The relation to proposed graviton-detection strategies is likewise
complementary. Carney, Domcke, and Rodd have emphasized that observing
individual graviton-like events or associated shot noise does not, by itself,
provide an unambiguous demonstration of the quantization of the gravitational
field~\cite{PhysRevD.109.044009}. The present framework does not rely on resolving
individual gravitons. Instead, it considers a correlation-based observable:
the generation of detector--detector entanglement through an explicitly
retained propagating gravitational degree of freedom. This provides a
different operational route for probing the quantum character of
gravitational radiation.

A related experimental direction has been explored using quantum acoustic
resonators, where discrete graviton--matter energy exchange can provide a
potential signature of quantum gravitational radiation~\cite{Tobar:2023ksi}.
Those proposals focus on discrete energy transfer between gravitational
radiation and a quantum mechanical system. The present framework addresses a
different question: whether a common propagating quantum gravitational field
can establish nonseparable correlations between separate quantum detector
subsystems. These approaches therefore probe distinct manifestations of the
quantum gravitational field and should be regarded as complementary rather
than competing strategies.

A particularly interesting extension concerns the interplay between gravity-induced entanglement and decoherence for more general initial states of the gravitational field. In our recent work \cite{Nandi:2026sww}, we showed that the reduced detector dynamics can depend qualitatively on the initial gravitational state. In particular, the vacuum exhibits a protected detector sector at leading order, whereas finite graviton occupation can lift this protection and introduce additional decoherence channels. This raises a natural question for the present entanglement-generating mechanism: how robust is the generated detector entanglement when the gravitational field is initially in a populated or otherwise non-vacuum state?

A systematic investigation would require comparing the characteristic entanglement-generation timescale, ($\tau_{\mathrm{ent}}$), with the gravitationally induced decoherence timescale, ($\tau_{\mathrm{dec}}$), using a consistent reduced-state description and an appropriate entanglement measure. In particular, it would be interesting to determine whether there exist parameter regimes in which entanglement develops before the relevant detector coherences are substantially degraded. Such an analysis would clarify the interplay between the quantum correlations mediated by gravitational radiation and the state-dependent decoherence induced by gravitational fluctuations.

There are, however, several limitations that should be kept explicit. The
gravitational field has been treated in the weak-field regime of linearized
General Relativity, and the detector--field interaction is considered
perturbatively. In addition, we retain a single gravitational mode as an
effective narrow-band description. This approximation is useful for
isolating the leading resonant dynamics, but it does not imply that
off-resonant modes are physically absent. In a complete multimode
description, off-resonant modes can contribute through virtual intermediate
processes and associated dispersive corrections. The present single-mode
model assumes that such cumulative corrections remain perturbatively small
on the dynamical scales relevant to the effective description.

The two-stage detector analysis should also be interpreted in this context.
The projected four-dimensional sector was introduced only to isolate the
lowest-order excitation pathways responsible for the emergence of detector
entanglement. Once this mechanism was established, the analysis was
returned to the complete six-dimensional detector subspace populated at the
perturbative order considered. The latter provides the appropriate setting
for characterizing the complete reduced detector dynamics, including the
entropy and purity. Thus, the projection is a diagnostic tool for identifying
the leading entanglement mechanism rather than a replacement for the full
perturbative detector state.

From an experimental perspective, the detector--graviton coupling predicted
within the present model is extremely weak. The results should therefore
not be interpreted as an immediate proposal for detecting individual
gravitons with existing gravitational-wave observatories. Rather, they
provide a theoretical framework for identifying quantum signatures of
propagating gravitational radiation using highly coherent quantum systems.
Future developments could investigate realistic environmental decoherence,
multimode gravitational fields, different initial gravitational states,
including coherent or non-Gaussian states, and extensions beyond the
linearized gravitational interaction.

\section{Conclusion}
\label{sec:conclusion}

We have developed an operational framework for investigating quantum
signatures of gravity in the dynamical regime of propagating gravitational
radiation. Two localized quantum detector systems are coupled locally to
the same propagating gravitational-wave field, while no direct
detector--detector interaction is introduced. By retaining the gravitational
field as an explicit dynamical degree of freedom, the framework allows the
classical stochastic and quantized descriptions of the same propagating
field to be compared within a common detector--field interaction.

Our central result is that a quantized propagating gravitational field can
establish genuine entanglement between the two detector systems. The
coherent excitation pathways generated by the local detector--field
interaction lead, after tracing over the gravitational degrees of freedom,
to a reduced detector state with nonzero entanglement negativity. The
corresponding classical stochastic description, in contrast, produces a
mixed but separable detector state: each classical realization gives a
factorized detector evolution, and stochastic averaging cannot generate
detector--detector entanglement.

The significance of this result lies in treating the propagating
gravitational-wave degree of freedom itself as the common dynamical
mediator. Rather than inferring its character from an effective
detector--detector interaction, the analysis retains the mediator
explicitly and examines the correlations generated between independent
quantum probes. Within the assumptions of the independent-mediator
framework, detector entanglement therefore provides an operational
distinction between the classical stochastic and quantized descriptions of
propagating gravitational radiation.

The result extends gravity-mediated quantum-correlation studies into the
radiative regime and provides a complementary correlation-based perspective
on the quantum character of gravitational radiation. The framework remains
within the weak-field, perturbative, and single-mode approximations adopted
here, and its extremely weak gravitational coupling presents a substantial
experimental challenge. Nevertheless, the analysis identifies a route in
which quantum signatures of gravitational radiation are sought through
nonseparable correlations between quantum systems rather than through
individual graviton detection alone. Future extensions to multimode
gravitational fields, environmental decoherence and dissipation, and more
general gravitational and detector states will be important for assessing
the robustness and experimental relevance of this signature.

\section*{Acknowledgments}

I gratefully acknowledge the support of the National Institute for Theoretical and Computational Sciences (NITheCS) through the Rector's Postdoctoral Fellowship Programme (RPFP). It is a pleasure to thank Bibhas Ranjan Majhi and Mainak Dutta for their fruitful collaboration, insightful discussions, and many valuable suggestions throughout this work. I am also grateful to Ms. Nandita Debnath for helpful correspondence and useful discussions. I would like to acknowledge the correspondence with Kazuhiro Yamamoto and Akira Matsumura of the Quantum and Spacetime Research Institute at Kyushu University, and thank them for their fruitful comments, valuable insights, kind invitation, and warm hospitality. Finally, I thank the organizers of the \emph{Quantum Universe 2025} workshop for providing a stimulating scientific environment and for their warm hospitality.  Finally, I sincerely thank the anonymous referee for their constructive comments and valuable suggestions, which helped improve the quality and clarity of this work.

\section*{Appendices}

\appendix

\section{Derivation of the Hamiltonian for graviton modes}
\label{Appendix A}

The analysis follows \cite{Parikh:2020fhy,Nandi:2024jyf}. We start from
linearized gravity about Minkowski spacetime,
\begin{equation}
g_{\mu\nu}=\eta_{\mu\nu}+h_{\mu\nu},
\qquad
|h_{\mu\nu}|\ll 1,
\end{equation}
and impose the transverse-traceless (TT) gauge,
\begin{equation}
h_{0\mu}=0,
\qquad
\partial^i h_{ij}=0,
\qquad
h^i{}_i=0.
\end{equation}
The quadratic Einstein-Hilbert action for the propagating TT degrees
of freedom is
\begin{equation}
S_{\rm grav}
=
-\frac{c^3}{64\pi G}
\int d^4x\,
(\partial_\alpha h_{ij})
(\partial^\alpha h^{ij}).
\label{eq:grav_action_appendix}
\end{equation}

To quantize the field, we consider a cubic box of volume $L^3$ and
expand the metric perturbation as
\begin{equation}
h_{ij}(t,\vec{x})
=
\sqrt{\frac{c^3}{\hbar G}}
\sum_{\vec{k},s}
q_{\vec{k},s}(t)
e^{i\vec{k}\cdot\vec{x}}
\epsilon^{(s)}_{ij}(\vec{k}),
\label{eq:mode_expansion_appendix}
\end{equation}
where
\begin{equation}
\vec{k}=\frac{2\pi}{L}\vec{n},
\qquad
\vec{n}\in\mathbb{Z}^3,
\end{equation}
and the polarization tensors satisfy
\begin{equation}
k^i\epsilon^{(s)}_{ij}=0,
\qquad
\delta^{ij}\epsilon^{(s)}_{ij}=0,
\qquad
\epsilon^{(s)}_{ij}\epsilon^{ij}_{(s')}
=
2\delta_{ss'}.
\end{equation}
The overall normalization in Eq.~\eqref{eq:mode_expansion_appendix}
is a choice of mode coordinate. Other canonical normalizations are
possible and lead to a corresponding rescaling of the mode variables
and oscillator coefficient, without changing the physical metric
perturbation or observable predictions.

Substituting Eq.~\eqref{eq:mode_expansion_appendix} into
Eq.~\eqref{eq:grav_action_appendix}, and using the orthogonality of the
Fourier modes, gives
\begin{equation}
S_{\rm grav}
=
\frac{c^2L^3}{32\pi G l_p^2}
\int dt
\sum_{\vec{k},s}
\left(
|\dot q_{\vec{k},s}|^2
-
\omega_k^2|q_{\vec{k},s}|^2
\right),
\label{eq:grav_action_modes}
\end{equation}
where
\begin{equation}
\omega_k=c|\vec{k}|,
\qquad
l_p^2=\frac{\hbar G}{c^3}.
\end{equation}
Equivalently, in natural units $c=\hbar=1$,
\begin{equation}
S_{\rm grav}
=
\frac{L^3}{32\pi G^2}
\int dt
\sum_{\vec{k},s}
\left(
|\dot q_{\vec{k},s}|^2
-
\omega_k^2|q_{\vec{k},s}|^2
\right).
\end{equation}

It is convenient to introduce the effective oscillator parameter
\begin{equation}
m_{\rm eff}
=
\frac{c^2L^3}{16\pi G l_p^2}
=
\frac{L^3}{16\pi\hbar G^2},
\label{eq:meff_appendix}
\end{equation}
so that each independent mode has the harmonic-oscillator form
\begin{equation}
S_{\rm grav}
=
\frac{1}{2}
\int dt
\sum_{\vec{k},s}
m_{\rm eff}
\left(
|\dot q_{\vec{k},s}|^2
-
\omega_k^2|q_{\vec{k},s}|^2
\right).
\end{equation}
The quantity $m_{\rm eff}$ is an effective oscillator parameter
associated with our choice of mode normalization and should not be
interpreted as a physical graviton mass. Its dependence on the box
volume arises from the normalization of the discrete Fourier modes.

The canonical momentum is
\begin{equation}
p_{\vec{k},s}
=
m_{\rm eff}\dot q_{\vec{k},s},
\end{equation}
and the corresponding Hamiltonian is
\begin{equation}
H_{\rm grav}
=
\sum_{\vec{k},s}
\left[
\frac{p_{\vec{k},s}^2}{2m_{\rm eff}}
+
\frac{1}{2}
m_{\rm eff}\omega_k^2q_{\vec{k},s}^2
\right].
\label{eq:multimode_hamiltonian_appendix}
\end{equation}
A further canonical rescaling,
\begin{equation}
Q_{\vec{k},s}
=
\sqrt{m_{\rm eff}}\,q_{\vec{k},s},
\qquad
P_{\vec{k},s}
=
\frac{p_{\vec{k},s}}{\sqrt{m_{\rm eff}}},
\end{equation}
puts each mode into the unit-mass oscillator form. Thus, the
appearance of $m_{\rm eff}$ is convention dependent, while the
physical metric perturbation and the consistently normalized
interaction remain unchanged.

For the single gravitational mode retained in the main text, we
consider a plus-polarized mode propagating along the $+\hat z$
direction, with
\begin{equation}
\omega_g=c|\vec{k}|.
\end{equation}
Suppressing the mode labels, its Hamiltonian becomes
\begin{equation}
H_G
=
\frac{p_+^2}{2m_{\rm eff}}
+
\frac{1}{2}m_{\rm eff}\omega_g^2q_+^2.
\label{eq:single_graviton_hamiltonian}
\end{equation}
Upon canonical quantization,
\begin{equation}
[\hat q_+,\hat p_+]=i\hbar,
\end{equation}
and introducing the creation and annihilation operators in the usual
way, we obtain
\begin{equation}
\hat H_G
=
\hbar\omega_g
\left(
\hat b_g^\dagger\hat b_g+\frac{1}{2}
\right).
\label{eq:graviton_hamiltonian_appendix}
\end{equation}
The interaction-picture mode coordinate is therefore
\begin{equation}
\hat q_+(t)
=
\sqrt{\frac{\hbar}{2m_{\rm eff}\omega_g}}
\left(
\hat b_g e^{-i\omega_gt}
+
\hat b_g^\dagger e^{i\omega_gt}
\right).
\label{eq:q_plus_operator}
\end{equation}

For the plus-polarized mode, the metric perturbation can be written as
\begin{equation}
h_{ij}
=
2\chi(t)\epsilon_+\sigma^3_{ij}
=
\sqrt{\frac{c^3}{\hbar G}}\,
q_+(t)e^{i\vec{k}\cdot\vec{x}}
\epsilon^+_{ij},
\label{eq:metric_plus_mode}
\end{equation}
which gives
\begin{equation}
\chi(t)
=
\frac{q_+(t)e^{i\vec{k}\cdot\vec{x}}}{2l_p}.
\end{equation}
At the detector position $z=0$, the corresponding operator is
\begin{equation}
\hat\chi(t)
=
\frac{1}{2l_p}
\sqrt{\frac{\hbar}{2m_{\rm eff}\omega_g}}
\left(
\hat b_g e^{-i\omega_gt}
+
\hat b_g^\dagger e^{i\omega_gt}
\right).
\end{equation}
Since the detector couples to
\begin{equation}
\hat\gamma(t)=\frac{1}{2}\dot{\hat\chi}(t),
\end{equation}
we obtain
\begin{equation}
\hat\gamma(t)
=
iC_\gamma
\left(
\hat b_g e^{-i\omega_gt}
-
\hat b_g^\dagger e^{i\omega_gt}
\right),
\label{eq:gamma_appendix}
\end{equation}
where
\begin{equation}
C_\gamma
=
-\sqrt{
\frac{\omega_gc\pi l_p^2}{2L^3}
}.
\label{eq:Cgamma_appendix}
\end{equation}
Thus, $C_\gamma$ has dimensions of frequency, as required by the
interaction Hamiltonian used in the main text. Its
$\sqrt{\omega_g}$ dependence follows from the derivative coupling:
the oscillator amplitude scales as $\omega_g^{-1/2}$, while taking the
time derivative contributes a factor of $\omega_g$. Hence the resulting
coupling scales as $C_\gamma\propto\sqrt{\omega_g}$ and does not
represent a frequency-dependent graviton mass.

%The projection removes the higher local excitation states
%$|04\rangle$ and $|40\rangle$, thereby isolating the sector that contains
%the lowest-order detector--detector correlations responsible for the
%emergence of bipartite entanglement. In the ordered basis

%\[
%\{|00\rangle,|02\rangle,|20\rangle,|22\rangle\},
%\]

%the projected density operator becomes

%%%%%%%%%%%%%%%%%%%%%%%%%%%%%%%%%%%%%%%%%%%%%%%%%%%%%%%%%%%%%%%

%\bibliographystyle{apsrev4-2}
%\bibliography{gw_phases}
%%%%%%%%%%%%%%%%%%%%%%%%%%%%%%%%%%%%%%%%%%%%%%%%%%%%%%%%%%%%%%%

\bibliographystyle{unsrt}

\bibliography{gw_phases}

%\bibliography{references}

%%%%%%%%%%%%%%%%%%%%%%%%%%%%%%%%%%%%%%%%%%%%%%%%%%%%%%%%%%%%%%%

\end{document}